\documentclass[aps,prfluids,superscriptaddress,onecolumn,longbibliography]{revtex4-2}

\usepackage{graphicx}
\usepackage{natbib}
\usepackage{amsmath}
\usepackage[version=4]{mhchem}
\usepackage{siunitx}
\usepackage{xfrac}% Unused?
\usepackage{physics}
\usepackage{placeins}
\usepackage{soul,color}
\newcommand{\recrit}{\Re_{\text{crit}}}
\begin{document}

\title{Vortex Breakdown in the Shear-Driven Flow in a Rectangular Cavity}

\author{Haoyi Wang}
\affiliation{Center for Fluid Mechanics, School of Engineering, Brown University, Providence, Rhode Island 02912, USA}

\author{Xinyi Yu}
\affiliation{Center for Fluid Mechanics, School of Engineering, Brown University, Providence, Rhode Island 02912, USA}

\author{San To Chan}
\affiliation{Okinawa Institute of Science and Technology Graduate University, Onna, Okinawa 904-0495, Japan}

\author{Guillaume Durey}
\affiliation{Center for Fluid Mechanics, School of Engineering, Brown University, Providence, Rhode Island 02912, USA}

\author{Amy Shen}
\affiliation{Okinawa Institute of Science and Technology Graduate University, Onna, Okinawa 904-0495, Japan}

\author{Jesse T. Ault}
\email{jesse\_ault@brown.edu}
\affiliation{Center for Fluid Mechanics, School of Engineering, Brown University, Providence, Rhode Island 02912, USA}
\date{\today}

\begin{abstract}
The vortex dynamics of laminar flow past a rectangular cavity is investigated using numerical simulations and microfluidic experiments. The flow is inherently three-dimensional and is characterized by a large, dominant vortex structure that fills most of the cavity at moderate Reynolds numbers with a weak, yet significant flow in the axial direction along the vortex core. Classical bubble-type vortex breakdown is observed within the cavity above a certain critical Reynolds number, which is a function of the channel width. The critical Reynolds number for the onset of breakdown is determined as a function of channel width, and the evolution and dynamical transitions of the breakdown regions are investigated as functions of the channel width and Reynolds number. At large cavity widths, two vortex breakdown bubbles emerge near the sidewalls symmetric about the centerplane, which grow and eventually merge as the Reynolds number increases. For large-enough widths, the vortex breakdown regions remain well-separated and their structures become independent of the cavity width. The stability and bifurcations of the stagnation points and their transitions to stable/unstable limit cycles are analyzed, and the criticality of the vortex flow is calculated, demonstrating that the vortex breakdown in the cavity agrees with Benjamin's interpretation of criticality. At the intermediate width regime, a single vortex breakdown bubble appears above the critical Reynolds number. In the narrow width regime, the flow exhibits more complicated modes. An additional vortex breakdown mode with reversed flow patterns is observed in this width regime, along with multiple shifts in the stability of stagnation points. The experimental and numerical results also demonstrate the sensitivity of the flow to the inlet conditions, such that relatively small asymmetries upstream can result in significant changes to the vortex breakdown behavior in the cavity.
\end{abstract}

\maketitle

\section{Introduction}

\subsection{Shear- and Lid-Driven Flow in a Cavity}

Despite the relatively simple geometry of a flow past a rectangular cavity, such flows can exhibit a surprisingly wide range of complex behaviors, such as corner eddies, non-unique flow solutions, longitudinal vortices, as well as different instabilities and types of transition \cite{ratkowsky1968viscous, cavity2000}. This complexity, along with the canonical geometry of such systems, has led these flows to receive a great deal of scientific attention and research. Much of the initial research into such systems focused on the simplified 2D version of the flow, considering both the lid-driven and shear-driven forms of the problem. These preliminary investigations revealed fairly intuitive dynamics wherein for either case the predominant flow structure in the cavity is a single large eddy structure that tends to occupy the majority of the cavity, with secondary eddies in the corners \cite{higdon1985stokes, ghia1982solution}. Studies that considered 3D effects showed that introducing a spanwise dimension and side walls can significantly change the qualitative behavior of such flows \cite{de1976evaluation}. For such flows at inertial but laminar Reynolds numbers, the 3D flow still remains dominated by a single main eddy structure, except that the spanwise velocity along the vortex core is not necessarily zero. Although this velocity component is small relative to the swirl velocity, it has previously been found to point from the channel centerplane towards the side walls, and such flows have been found to exhibit symmetry about the channel centerplane \cite{chiang1996finite}. This axial flow from the centerplane towards the wall along the vortex core indicates the presence of a stagnation point at the channel center on the vortex core, which is unstable on the vortex axis and stable on the centerplane. Here, when we refer to the stability of stagnation points, we are simply indicating the direction of flow either toward or away from the stagnation point. That is to say, if we refer to a stagnation point as unstable on the vortex axis, this implies that a tracer particle located at the stagnation point would flow away from the stagnation point along the axis if perturbed in that direction and vice versa. The combination of a stagnation zone, strong swirl, and weak axial flow along the vortex core hints at the potential for the onset of vortex breakdown in such systems, though most previous studies have ignored such effects. Several other key studies that have investigated 3D cavity flow have considered effects such as the influence of end walls on the formation of Taylor-type instabilities and Taylor-G{\"o}rtler-like vortices in the lid-driven system \cite{koseff1984end, koseff1984lid, koseff1984visualization}. Related research for the shear-driven system examined the dynamics of the shear layer between the outer channel flow and the flow in the cavity showing that systems with either deep or shallow cavity depths can become unsteady at high Reynolds numbers via Kelvin-Helmholtz instability oscillations \cite{yao2004numerical}.

One specific application of shear-driven cavity flows that has emerged in recent years is in the context of drag reduction. Specifically, researchers have used patterned or porous surfaces to reduce the laminar or turbulent drag over a surface by infusing the surface with various fluids such as simply air or a variety of oils \cite{solomon2014drag, wexler2015shear, wexler2015robust, liu2016effect, rosenberg2016turbulent, alinovi2018apparent}. By doing this, the boundary condition of the external flow at the surface acquires some effective slip length that reduces the drag overall. The simplest example of this to consider is the shear-driven flow past a single cavity or array of cavities. As the external flow goes past this patterned surface, it encounters alternating no-slip conditions with the surface where there is no cavity, and effective slip conditions where the cavities are present. By infusing such cavities with different fluids such as oils, as well as alternating the geometry of the cavity, various drag reduction efficiencies can be achieved. Researchers have investigated such systems with various approaches. For example, \citeauthor{wexler2015shear} studied the shear-driven failure of such surfaces when the cavities were infused with different oils and investigated approaches for reducing this failure mechanism \cite{wexler2015shear}. \citeauthor{wexler2015robust} also used chemical patterning strategies to change the preferential wetting of the cavities in an attempt to prevent this leakage mechanism \cite{wexler2015robust}. Further research investigated the role of the viscosity ratio on the shear-driven failure mode and showed that lower viscosity of the driving fluid caused less fluid draining \cite{liu2016effect}. \citeauthor{solomon2014drag} systematically demonstrated the ability of such surfaces to reduce drag in a robust way for laminar flows, developing scaling models to determine the dependence of drag reduction on the viscosity ratio \cite{solomon2014drag}. \citeauthor{rosenberg2016turbulent} used fully turbulent experiments to demonstrate up to a 14\% drag reduction using such liquid-infused patterned surfaces in a Taylor-Couette flow \cite{rosenberg2016turbulent}. Finally, \citeauthor{alinovi2018apparent} decoupled the turbulent flow past such a patterned surface into the small-scale Stokes problem within the cavity and coupled this to numerical simulations of the outer turbulent flow problem, further characterizing the efficiency of such drag-reduction strategies in turbulent flows \cite{alinovi2018apparent}.

\begin{figure}
    \centering
    \includegraphics[width=0.7\textwidth]{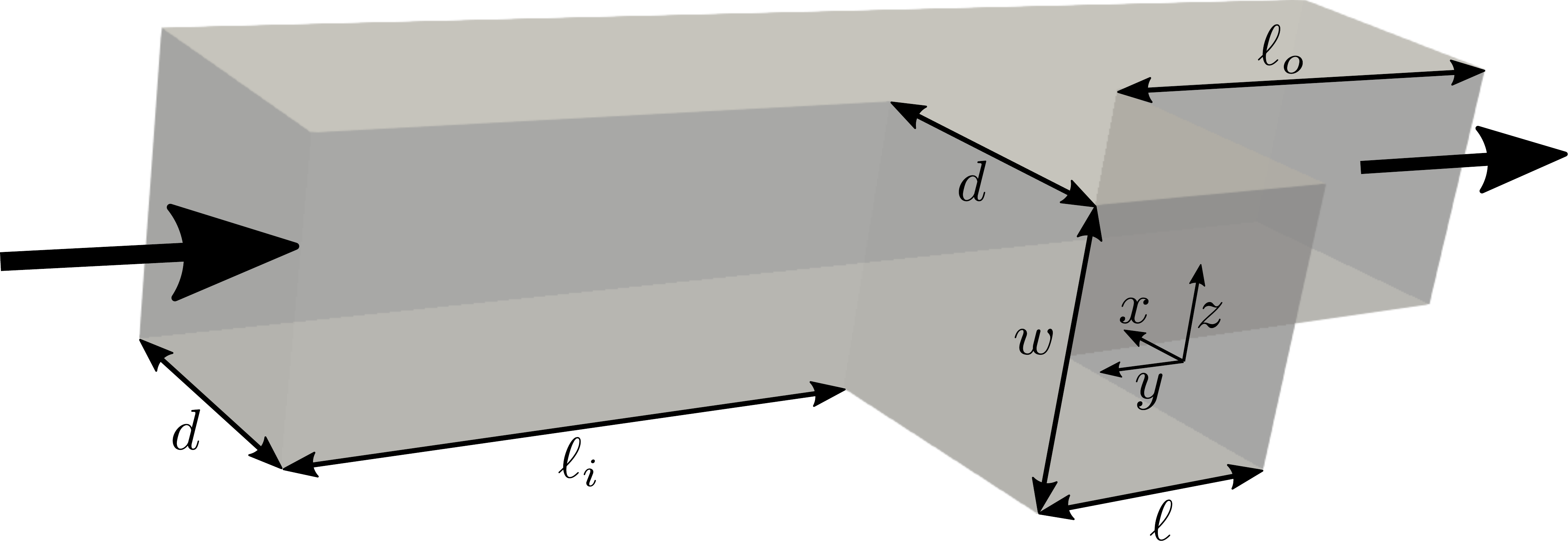}
    \caption{System setup and coordinate system for studying the shear-driven flow past a rectangular cavity.}
    \label{fig:geometry}
\end{figure}

In the current work, we show that the shear-driven flow in a cavity can exhibit vortex breakdown, a phenomenon that can potentially alter the mass transport, flow stability, and slip behaviors in systems such as those described above. An illustration of the specific system we study is shown in Figure \ref{fig:geometry}.
The system consists of a channel flow past a single rectangular cavity of length $\ell$, width $w$, and depth $d$. The inlet and outlet lengths of the channel are $\ell_i$ and $\ell_o$, respectively. To reduce the potential parameter space, we fix the main flow channel to be square with side length $d$, and we use sufficiently long inlets and outlets such that the fluid dynamics in the cavity are independent of $\ell_i$ and $\ell_o$. With these constraints, the two key nondimensional numbers that fix the geometry of our system are $w/d$ and $\ell/d$. For this preliminary investigation, we also consider cases where $\ell/d=1$ to further constrain the geometry. We define $\Re=\frac{U\ell}{\nu}$, where $U$ is the average inlet velocity and $\nu$ is the kinematic viscosity of the fluid. For simplicity, we place the origin of the geometry at the center of the bottom wall. The centerplane is therefore $z=0$. At any point along the vortex core, we characterize the vortex as having an \textit{outward axial flow} if it is directed away from $z=0$, and an \textit{inward axial flow} if it is directed toward $z=0$. Before proceeding to describe our methodology and characterize the types of flow behaviors we observe in such systems, we first provide a brief introduction to the vortex breakdown phenomenon in the following section.

\subsection{Vortex Breakdown}\label{vbintro}

In many vortex-dominated flows, the vortex core can often be approximated as axisymmetric. Within the vortex core, the balance of the axial and swirl velocity components plays an important role in determining the dynamics of the vortex. In particular, in certain flows the vortex core can experience a sudden transition in which an internal stagnation point appears in the flow. As the flow decelerates towards this point, the streamsurfaces suddenly diverge. This phenomenon is called vortex breakdown. Different types of such behaviors have been observed, the two most common of which are the spiral type and the nearly-axisymmetric bubble type breakdown \cite{hall1972vortex, sarpkaya1971vortex, lucca2001vortex, leibovich1978structure}. Early research into vortex breakdown primarily centered on the field of aeronautics \cite{wentz1971vortex, lowson1995vortex, peckham1960preliminary, rusak1999prediction}. For example, such breakdown behaviors were originally discovered in the relatively high-speed flows over delta wings with highly swept leading edges at steep angles of attack \cite{peckham1960preliminary, hummel1967vortex, gad1985discrete}. However, modern research has identified vortex breakdown in a much wider range of flow systems, from high-speed external flows to simple swirling pipe flows and even in low- to moderate-Reynolds number microfluidics \cite{ault2016vortex, oettinger2018invisible, shin2015flow, chan2019coupling, chen2017vortex, chan2018microscopic, chen2015vortex, peikert2007visualization, rusak1999prediction, rusak2012global, rusak1996review, harvey1962some, cassidy1970observations, sarpkaya1971stationary}.

Typically, theoretical studies of vortex breakdown have been performed in relatively basic geometries such as an axisymmetric pipe flow where the source of swirl may be either the rotating pipe wall, a rotating end wall in a closed pipe system, or an imposed theoretical inlet condition such as imposed Batchelor vortices, etc. For example, several studies have considered the vortex breakdown in an axisymmetric cylinder geometry with a rotating end wall. In particular, the three-part study on axisymmetric vortex breakdown by \citet{lopez1990axisymmetric}, \citet{brown1990axisymmetric}, and \citet{lopez1992axisymmetric} performed a detailed experimental, numerical, and theoretical investigation of the vortex breakdown dynamics in laminar swirling flows driven by a rotating end wall. Among other things, they identified the onset of oscillatory flows with simulations of the full unsteady axisymmetric Navier-Stokes equations which could not be well-resolved experimentally, they derived an inviscid equation of motion that lead to a theoretical criterion for vortex breakdown based on the generation of negative azimuthal vorticity on streamsurfaces, and they used nonlinear dynamical systems theory to investigate the dynamics of the flow across many rotation speeds, observing multiple distinct oscillation modes and chaotic advection. Later, \citet{marques2001precessing} performed additional simulations of this system to resolve an apparent discrepancy in the prior observations of the first mode of instability. Recently, \citet{li2022internal} performed simulations of a similar system, but across a wide range of cylinder aspect ratios and Reynolds numbers and observed stair-stepping changes in the locations of the vortex breakdown regions as the parameters were varied. In a different system, \citet{dennis2014controlling} considered the flow in an axisymmetric pipe with no end walls, where a portion of the pipe itself rotates to generate the swirl. In this way, they could consider both decaying and growing swirl in the flow direction, observing traditional bubble breakdown on the pipe axis for decaying swirl, and toroidal recirculation zones near the walls for growing swirl. Vortex dynamics and symmetry-breaking bifurcations have also been investigated in a range of other systems, such as in a cross-slot flow mixer, where above a certain Reynolds number, vortex stretching at the stagnation point drives instability that triggers the development of spiral vortex structures \cite{haward2016tricritical}.
 
 A variety of key studies have used theoretical and numerical techniques to explore the stability characteristics, path to transition, and control strategies for key systems exhibiting vortex breakdown. Some of this work has been done in the high-$\Re$ limit. For example, \citet{wang1996stability} studied the linear stability of an inviscid, axisymmetric swirling flow in a finite length pipe by deriving a set of linearized equations for axially symmetric perturbations imposed on a base columnar flow. This work revealed a novel instability mechanism for swirling flows that cannot be predicted from previous analysis and developed new understanding of the relationship between stability and the vortex breakdown phenomenon. \citet{gallaire2004closed} considered this same finite length pipe model and developed a closed-loop control scheme that can preserve the columnar swirling base state to prevent the linear development of the instability identified by \citet{wang1996stability}. In another work, \citet{wang1997dynamics} used the axisymmetric unsteady Euler equations to investigate the stability characteristics and time-asymptotic behavior of an inviscid axisymmetric swirling flow and developed a consistent explanation for the mechanism that leads to vortex breakdown in high-$\Re$ swirling flows in a pipe. \citet{gallaire2004role} followed this same analysis to further consider how perturbations destabilize a columnar swirling jet, and they related the local stability properties in an infinite pipe to the global stability properties in a finite pipe.
 
 Despite the classic nature of vortex breakdown in the fluid dynamics community, controversies about its physical origins remain. Various explanations have been explored, but none of these have received broad acceptance in the community \cite{hall1972vortex, spall1987criterion, keller1995interpretation, darmofal1994trapped, pasche2018predictive, escudier1983vortex}. The three main categories for these explanations of vortex breakdown are summarized by Hall \cite{hall1972vortex} as: (1) Vortex breakdown is analogous to 2D boundary-layer separation \cite{gartshore1962recent}, (2) Vortex breakdown is due to hydrodynamic instability \cite{ludwig1962}, and (3) Vortex breakdown depends on the existence of a critical state in which the vortex can support certain types of standing waves \cite{squire1960analysis, bossel1967inviscid, benjamin1962, benjamin1967some}. Theoretical progress has been made toward each of these explanations, but the third appears to be the most favorable. Towards this explanation, Ruith \textit{et al.}, for example, has proposed a method to predict the criticality of vortex breakdown regions by axisymmetrizing the vortex core and then seeking zeros to the solutions of a particular differential equation within the axisymmetrized region \cite{ruith2003three}. One outstanding issue is that such types of analysis still require a priori knowledge of the flow field and the approximate location of any potential vortex breakdown regions. Recent work by \citeauthor{katsanoulis2023approximate} reported an algorithm to find the closest first integral to a given vector field and construct approximate streamsurfaces for the flow, facilitating the detection of vortex breakdown regions. Nonetheless, it still requires the full simulated flow field \cite{katsanoulis2023approximate}.

Furthermore, although vortex breakdown has typically been considered in the context of single-phase flows, recent work shows that it can have significant consequences for mass transport and mixing in multi-phase and particle-laden flows. For example, while investigating particle-wall impacts in a branching T-junction flow, Vigolo \textit{et al.} demonstrated unexpected bubble trapping that they later attributed to the onset of vortex breakdown above a critical Reynolds number of $\Re\approx320$, where there $\Re$ is based on the side dimension of the square inlet channel \cite{vigolo2013experimental, vigolo2014unexpected}. \citeauthor{ault2016vortex} later expanded on this work by considering the role of both the geometry and strength of the flow through a branching junction in triggering vortex breakdown \cite{ault2016vortex, chen2017high}. They combined single-phase simulations and bubble- and particle-laden microfluidics experiments to show that vortex breakdown in the single-phase flow is a necessary condition for the onset of particle/bubble capture in the experiments. However, it is not a sufficient condition, since particles and bubbles also experience centrifugal effects as they swirl around the vortex core, such that only low-density objects will be pulled toward the core and ultimately trapped. Later, \citeauthor{oettinger2018invisible} used numerical simulations along with an analytical approach to show that this capture is due to anchor-shaped 3D flow structures, and they identified the particle size and density regimes for capture via a stability analysis \cite{oettinger2018invisible}. \citeauthor{shin2015flow} demonstrated a practical application of this system by demonstrating the capture and rapid fusion of lipid vesicles in the high-shear vortex breakdown regions as a novel system to produce Giant Unilamellar Vesicles (GUVs) \cite{shin2015flow}. A similar system was used by introducing a small cavity in the vicinity of the junction to develop a versatile intracellular delivery method \cite{hur2020microfluidic}. Beyond these examples, vortex breakdown in general can have broad applications across microfluidics for all areas of mass transport, by introducing symmetry breaking, unsteadiness, unexpected particle/bubble trapping, or even local hot spots in reacting flows \cite{zhang2020trapping, lee2020three}. Thus, it is critical to be able to understand and predict when vortex breakdown will occur in such systems and how the breakdown will affect dynamics such as particle/mass transport and mixing. For a more comprehensive overview of the nonlinear and inertial fluid dynamics in microfluidic systems including those described above, see \cite{stoecklein2018nonlinear}.

Upon first consideration, the shear-driven flow in a cavity does not seem to be relevant to vortex breakdown. In particular, the natural first consideration of such a problem as a 2D system precludes the possibility of breakdown since no axial flow along the vortex core is possible. But even in the 3D case, breakdown does not at first appear to be likely. At least qualitatively the system is quite different from the typical systems that can experience breakdown. Typically, such flows have a strong axial flow component along the vortex core and correspondingly strong swirl velocities. Furthermore, the vortex axis is typically parallel or at least nearly aligned with the incoming main flow (e.g. the flow through a rotating pipe). In contrast, in the shear-driven flow in a cavity, the vortex axis is perpendicular to the incoming flow. The component of flow along the vortex core is quite small, going to zero in the limit of small cavity widths $w$. Nonetheless, we will show that cavity flow can experience multiple types of bubble-type vortex breakdown that evolve and transform as the cavity geometry and Reynolds numbers are varied.

\begin{figure}
    \centering
    \includegraphics[width=0.48\textwidth]{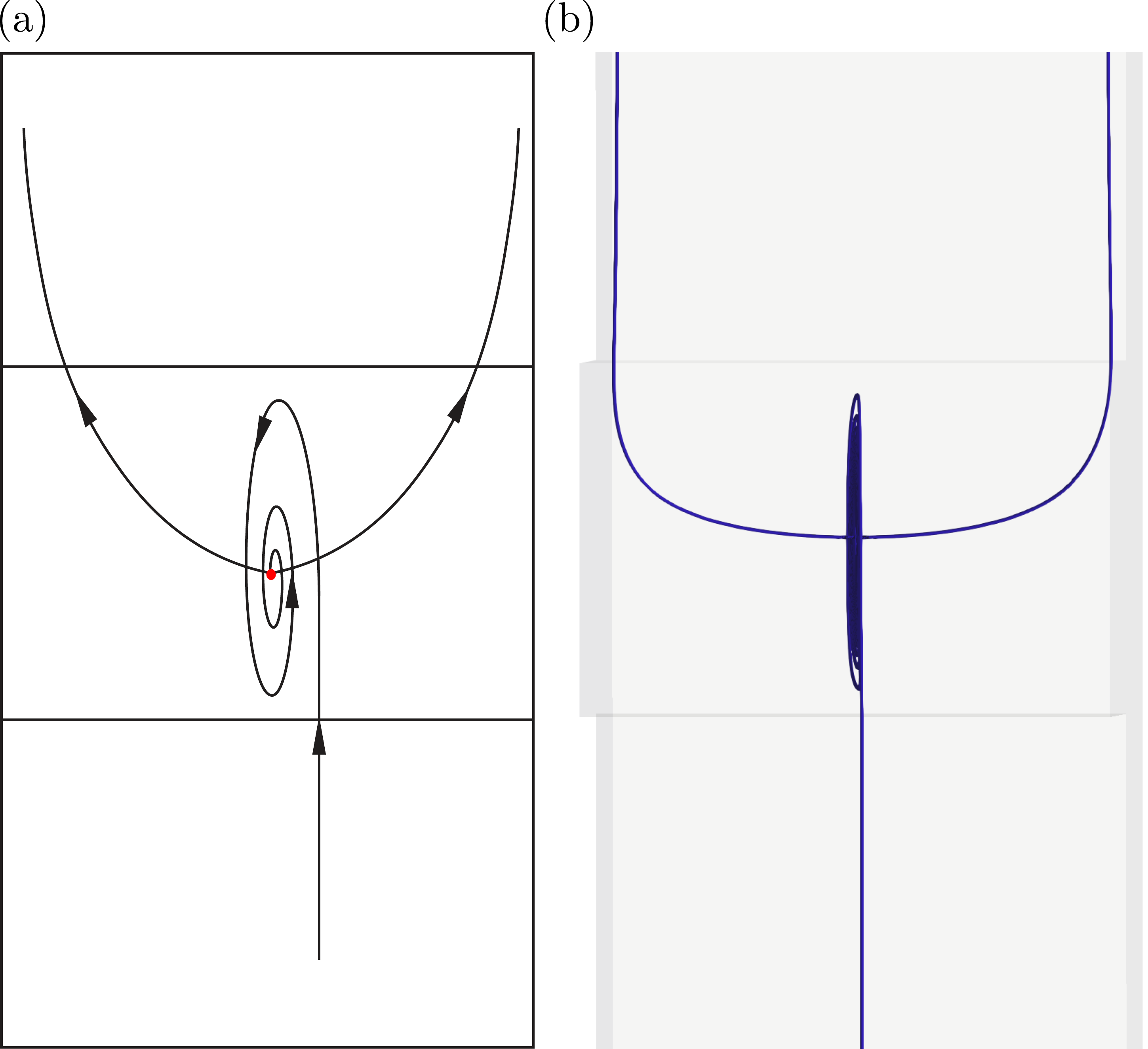}
    \includegraphics[width=0.48\textwidth]{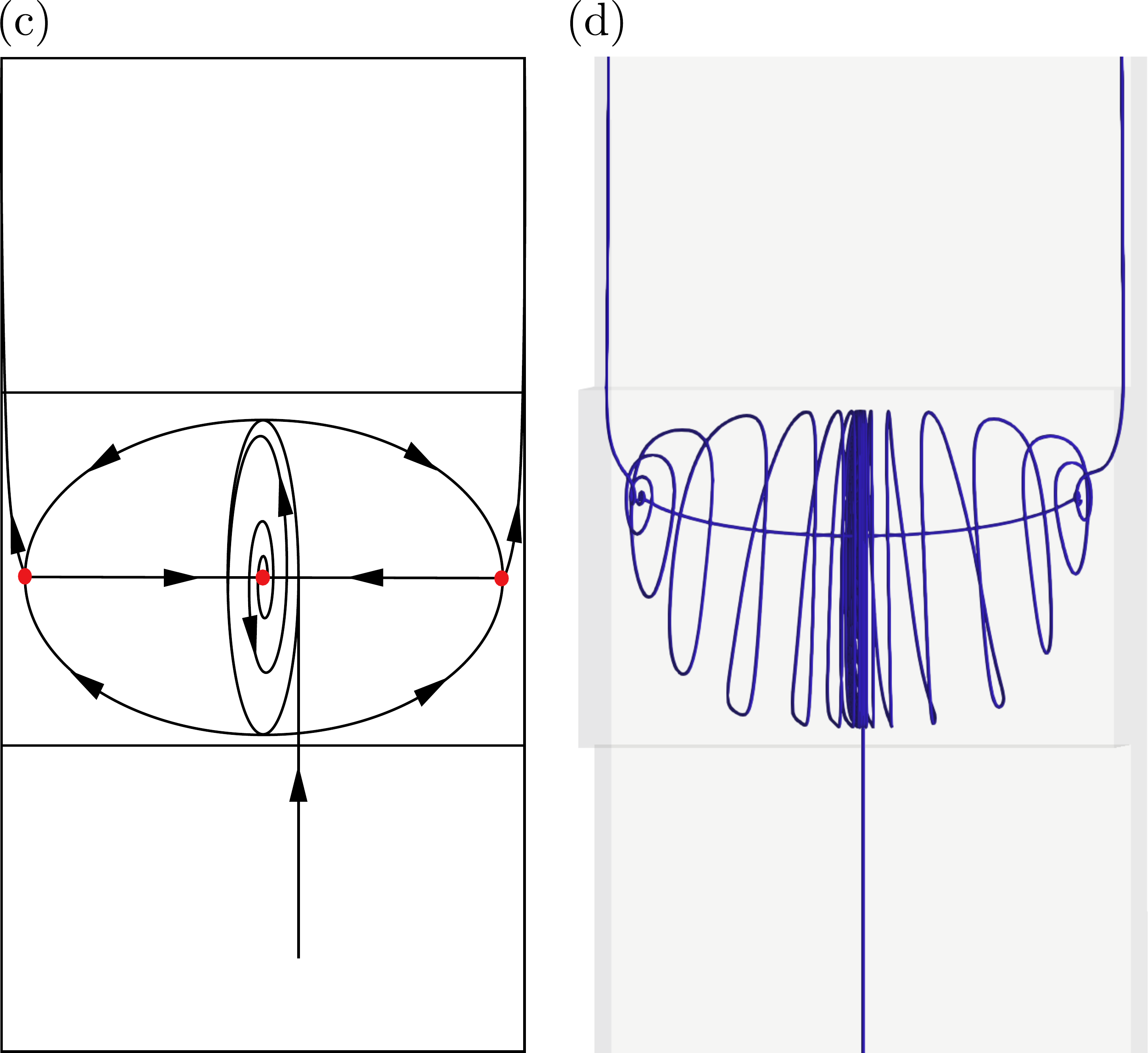}
    \caption{Schematics illustrating the two fluid flow behaviors described by \cite{torczynski} at Reynolds numbers of 35 (a,b) and 100 (c,d). Panels (a) and (c) correspond to the diagrammatic depictions of the qualitative explanation of the dynamics by \cite{torczynski} at the two Reynolds numbers, and panels (b) and (d) correspond to individual streamline visualizations from our 3D numerical simulations at similar flow conditions. Stagnation points are indicated by red dots. The flow clearly goes through a dynamical transition at some intermediate Reynolds number in which two new stagnation points appear and bound a new closed recirculation region that closely resembles classical bubble-type vortex breakdown. Here, the flow is from bottom to the top.}
    \label{fig:torczdemo}
\end{figure}

Studies on inertial microfluidic cavity flow have previously demonstrated particle trapping mechanisms within the cavity. For example, \citeauthor{haddadi2017inertial} studied the particle-laden flow over a narrow, long cavity ($w/\ell=0.22$, $\ell/d=2-5$) and observed recirculating, closed particle trajectories within the cavity \cite{haddadi2017inertial}. Trapping has also been observed in width-confined flows over different cavity geometries such as in cavities that are symmetric on both sides of the main flow channel \cite{shen2019experimental} and in round cavities \cite{shen2021round}. While vortex breakdown has previously been explored as a mechanism for particle trapping, it was not considered in the context of these studies. Specifically, these studies focused on width-confined channels ($w/\ell <0.25$ in all cases), which precludes the possibility of vortex breakdown due to the nearly negligible flow along the axial direction of the vortex. Specific evidence for the possibility of vortex breakdown in cavity flows was demonstrated in the numerical simulations of \citeauthor{torczynski}, in which they showed two distinct flow regimes as illustrated in Figure \ref{fig:torczdemo} for Reynolds numbers of 35 and 100.

Here, panels (a) and (c) correspond to diagrammatic depictions of the flow features seen by \citeauthor{torczynski} at the two Reynolds numbers, and panels (b) and (d) correspond to individual streamline visualizations from our numerical simulations for similar conditions. At $\Re=35$, \citeauthor{torczynski} observed that the incoming flow along the centerline spirals down into the cavity toward a stagnation point, where it then diverges off of the centerplane in two streamlines that proceed out of the cavity and downstream. This is consistent with the idea of a dominant eddy structure occupying the cavity with a relatively weak outward axial flow along its axis. For $\Re=100$, \citeauthor{torczynski} observed a `kidney bean shaped' closed region in which two additional stagnation points appeared near the side walls symmetric about the centerplane. These stagnation points appeared to bound the bubble region, defining the limits of a recirculation zone. These two new stagnation points are unstable along the vortex core and stable along the surface of the bubble region.

In addition, \citeauthor{torczynski} also report that the stagnation point on the centerplane experiences a shift in its stability, becoming unstable on the centerplane and stable on the vortex axis at this higher Reynolds number. Fluid along the surface of this closed region spirals toward the bounding stagnation points, where it then diverges back toward the centerplane along the vortex axis, and then spirals back out toward the bounding surface. Thus, this region represents a 3D closed recirculation region that resembles the classical bubble-type vortex breakdown. In this study, we expand on these preliminary findings by \cite{torczynski} by characterizing the vortex breakdown dynamics in such systems across a wide range of cavity geometries and Reynolds numbers. In particular, we highlight multiple transitions between different breakdown types using a combination of numerical simulations and microfluidics experiments. We perform criticality calculations on the identified recirculation zones to show that they demonstrate criticality consistently as expected for vortex breakdown regions. We also demonstrate the strong sensitivity of the vortex breakdown structure to relatively weak asymmetries in the incoming flow. To briefly outline the remaining structure of this paper, in \S\ref{section:methodology}, we describe our experimental and numerical methods, in \S\ref{section:numerical} we present both our numerical and experimental results, along with analysis and discussion. In \S\ref{section:conclusions} we present conclusions.

\section{Methodology}\label{section:methodology}

\subsection{Numerical Methods}\label{section:numerical_methods}

Fully three-dimensional computational fluid dynamics simulations were performed using the open source toolkit OpenFOAM \cite{openfoam}. Steady-state solutions were obtained using the \texttt{simpleFoam} solver that uses the Semi-Implicit Method for Pressure Linked Equations (SIMPLE) algorithm \cite{patankar2018numerical}. We verified that the steady-state solutions were stable by using the \texttt{icoFoam} transient flow solver that is based on the Pressure-Implicit with Splitting of Operators (PISO) method \cite{issa1986solution, issa1986computation}. Simulation geometries were created using the built-in mesh-generation tool \texttt{blockMesh}, with total numbers of cells in the domains ranging from $4.5\times 10^6$ for the smallest width cases with $w/\ell\le1$ up to $1.125\times 10^7$ for the largest width case with $w/\ell=15$. The simulation domains consist of hexahedral cells that are uniform in length in each of the three dimensions, ranging from side lengths of $\ell/70$ to $\ell/50$. For each case, convergence tests were performed with simulations containing both one-half and double the number of cells to quantify the uncertainties in the calculated critical Reynolds numbers for the onset of vortex breakdown. By comparing the critical $\Re$ value where vortex breakdown first occurs using each of these three grid resolutions (0.5x, 1x, and 2x), we found that the critical transition Reynolds number for the 1x grid differs by at most 3\% from the 2x grid. Individual differences are measured for each case and presented as error bars in the Results section.

Unless otherwise specified, the inlet and outlet lengths for each case were fixed at $2\ell$. The steady-state simulations converged to velocity and pressure residual tolerances of $1\times10^{-6}$ in approximately 0.5 hours when running on 32 cores on the Brown University Center for Computation and Visualization high performance computing cluster. The boundary conditions for the simulations were specified as follows: No-slip and zero normal pressure gradient conditions were imposed on all of the walls. At the channel inlet, unless specified otherwise, the velocity profile was set to be the analytical solution for the fully developed rectangular channel flow. Furthermore, a zero normal pressure gradient condition was set at the inlet, which improves stability and convergence of the solver, but introduces some minor errors within the first couple of cells of the inlet. At the outlet, the fluid velocity is assumed to be fully developed, and so a zero normal velocity gradient condition was imposed, and the pressure was set to a uniform value of zero. Initial conditions were set to zero velocity and pressure everywhere. For the majority of cavity widths, steady-state simulations were performed with $\Re$ ranging from 20 to 250, with Reynolds numbers reaching up to 450 for widths around $w/\ell=5$ in order to resolve a specific transition that occurs in that regime which will be described below. The specific cases that were tested for stability were those with ($w/\ell$, $\Re$) values of (0.5, 40), (0.5, 70), (3, 80), (5, 130), and (5, 450), respectively. For these simulations, the converged steady-state results of \texttt{simpleFoam} were used as the initial condition in the transient \texttt{icoFoam} solver, and all cases were confirmed to be stable. Here, adaptive time-stepping was used to ensure a maximum Courant number less than 0.5.

Post-processing, visualizations, and streamline calculations were performed using the open-source data analysis software ParaView. Based on our prior experience, vortex breakdown regions were identified by judicious placement of individual streamlines. For this particular flow geometry, all of the symmetric flows have a vortex core that intersects a stagnation point on the centerplane, provided the Reynolds number is not too small. Thus, the flow along the vortex core and the vortex breakdown regions themselves can be quickly identified by seeding a single streamline on the centerplane ($z=0$) and adjusting its $x$,$y$ position to find this stagnation point. This process is helped by the fact that these streamlines on the centerplane spiral in or out from this stagnation point, which makes visual identification of its position straightforward.

As an additional check to validate the recirculation zones we identify in light of theoretical interpretations of vortex breakdown, we will also perform criticality calculations of the swirling flows in the cavity for select cases. As mentioned above, one of the key explanations for the vortex breakdown phenomenon is that it depends on the existence of a critical state. In particular, this was the interpretation given by \citeauthor{benjamin1962}, who analyzed axisymmetric swirling flow in a circular pipe of radius $R$ and argues that breakdown is a transition between subcritical regions, which are capable of supporting standing waves, and supercritical regions, which are not capable of supporting standing waves \cite{benjamin1962}. In practice, this criticality condition can be determined from a velocity field by solving the ordinary differential equation
\begin{equation}\label{eq:criticality}
\left[ \frac{\partial^2}{\partial r^2}-\frac{ \partial}{r \partial r}+\frac{1}{ru_z}\frac{\partial u_z}{\partial r}+\frac{1}{r^3u_z^2}\frac{\partial (ru_\theta)^2}{\partial r}-\frac{1}{u_z}\frac{\partial^2 u_z}{\partial r^2} \right] \phi_c=0,
\end{equation}
subject to the boundary conditions that $\phi_c(0)=0$ and $\phi_c'=1$ at either $r=0$ or $r=R$. If the solution $\phi_c(r)$ has a root (that we designate as $r_\text{crit}$) between $0<r<R$, then the region is subcritical. Otherwise the flow is supercritical. We will use this method below to verify whether the recirculation regions that we identify satisfy this existing interpretation of vortex breakdown. One complication of applying this theory to our numerical results is that it was strictly developed in the context of axisymmetric flows in a pipe with a clearly defined radius $R$. For our case, the flow inside the cavity is clearly not axisymmetric about the vortex core for the majority of cases, so we reserve this approach for select cases that can be reasonably axisymmetrized. Furthermore, the selection of the outer limit for this axisymmetrized region (equivalent to $R$ in Benjamin's work) is not obvious, so we tested a variety of limits. Since the vortex core within the cavity is not perfectly straight, we cannot simply define a cylinder of radius $R$ about the axis. Instead, at each location along the vortex core, we choose a circular region of radius $R=\ell/5$ that is perpendicular to the vortex core and we calculate the axisymmetrized azimuthal and swirl velocities on the circle by averaging in the azimuthal direction. We then solve equation (\ref{eq:criticality}) on the slice in order to determine whether the solution has a root within the domain. We then plot the first root, designated as $r_\text{crit}$, versus the position along the vortex core and compare this with our streamline visualizations generated from Paraview.

\subsection{Experimental Methods}

\subsubsection{Device fabrication}

Microfluidic experiments were performed for channels with nondimensional cavity widths $w/\ell$ of 0.55, 1.01, 1.48, 1.97, 2.84, 3.96 and 4.83. All but the largest width case were fabricated directly in glass using the LightFab 3D printer (LightFab GmbH, Germany) based on the selective laser-induced etching (SLE) technology \cite{burshtein20193d, hnatovsky2006fabrication}. In this process, the designed channel structures are printed in a rigid piece of fused silica using ultrafast laser pulses which alter the chemical structure of the glass. The printed structures are then chemically etched with potassium hydroxide in an 80 $^\circ$C ultrasonic bath. The results are multiple transparent, rigid pieces of glass with the microfluidic cavities/channels embedded. Compared to the more commonly used polydimethylsiloxane (PDMS) or polymethyl methacrylate devices, the glass devices can endure higher pressure, making them more suitable for the moderate and high $\Re$ experiments conducted in the current study. In dimensional units, the fabricated microfluidic cavities have widths and depths in mm of $(w,~\ell)$ = (0.47, 0.85), (0.89, 0.88), (1.29, 0.87), (1.68, 0.85), (2.50, 0.88), and (3.30, 0.83), respectively.

The inlet and outlet lengths for all cases were designed to be $9.5w$ to ensure the flow is fully developed before reaching the cavity. Due to limitations in the size of devices that can be fabricated, SLE was not feasible to produce devices with $w/\ell>4$ with sufficiently long inlets and outlets to achieve fully developed flow before the cavity. However, as we will show later in the Results section, one of the key vortex breakdown transitions that we identified occurs in the range $4<w/\ell<5$. Thus, we used soft-lithography to create one more flow channel with $w/\ell=4.83$. To produce this channel, we first milled out a model of the channel from raw aluminum using an end mill. For this case, the milled dimensions of the mold were $w$ = 2.07 mm, $\ell$ = 10.0 mm, $\ell_i$ = 7$w$, and $\ell_o$ = 2$w$. This model was then glued to a glass slide, and Sylgard 184 PDMS was then used to create the channel from the mold. The PDMS channel was then bonded to a clean piece of glass after plasma treatment to ensure a secure bond.

\subsubsection{Experimental procedures}

\begin{figure}
    \centering
    \includegraphics[width=\textwidth]{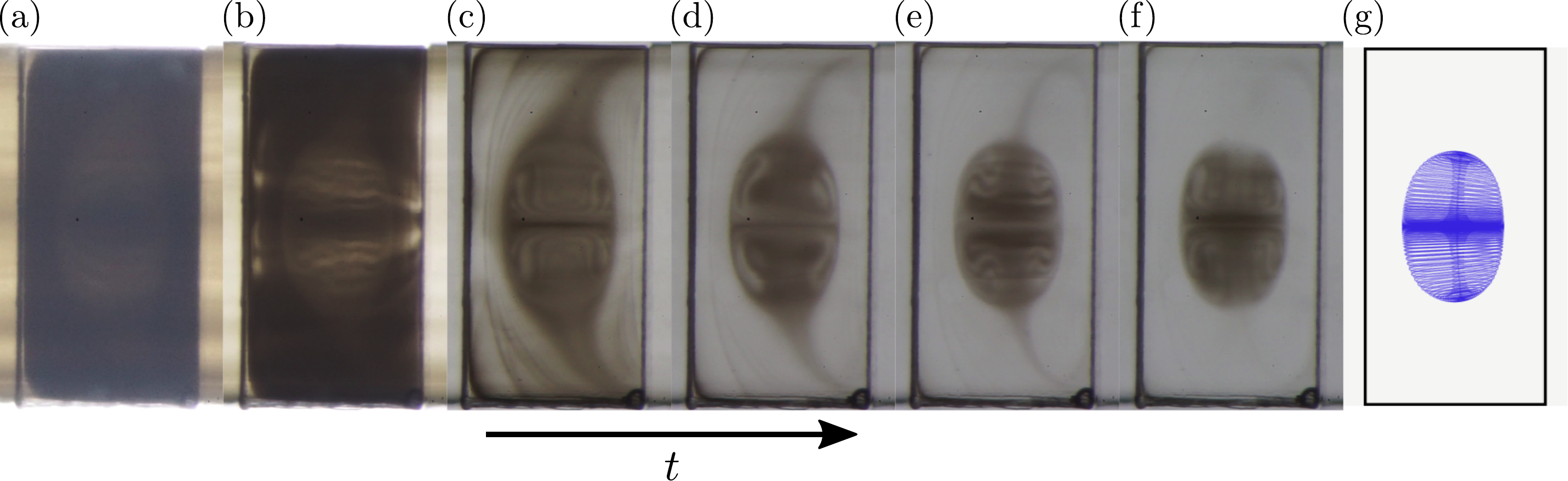}
    \caption{Bottom-view time-lapse visualization of the dye trapping experiments for a channel with $w/\ell=1.97$ at a Reynolds number of $\Re=59.6$. In the experiments, a dye-filled channel is flushed with fresh water, making the vortex breakdown recirculation bubble visually identifiable. Panels (a)-(f) are captured at a time interval of 2 s. Results are compared to a streamline visualization of the single-phase simulations in panel (g) from the numerical results with $w/\ell=2$ and $\Re=60$. The flow is from left to right.}
    \label{fig:timelapse}
\end{figure}

Vortex breakdown structures are visualized using a dye trapping approach that is based on the recirculating nature of the vortex breakdown regions. The flow channels are initially flushed with a dye made from diluted black watercolor ink until the entire channel and the cavity are saturated by dye. Next, clear water is flushed through the channel at a constant flow rate to set the desired Reynolds number of the experiment. The fresh water flushes away the dye except inside the vortex breakdown regions, where the streamlines form closed recirculation zones. These regions slowly fade as the dye diffuses out of the channel. To control the flow, we utilize a Harvard Apparatus PHD ULTRA syringe pump using a 60 mL syringe for the largest width case and a neMESYS (Cetoni GmbH, Germany) syringe pump with a 25 mL syringe for all of the other cases. The syringe pumps accelerate rapidly and achieve the target flow rates in less than 0.5 seconds. Experimental videos recording the flow behavior within the cavity were captured using a Leica K5 camera with a Leica DMi8 microscope for the largest width case and a SONY ILCE-6000 digital camera with a Nikon SMZ1270 stereo microscope for all other cases. For most cases with widths $w/\ell\geq 1$ (e.g. Figure \ref{fig:timelapse}),
this visual identification process is rather straightforward and provides reasonable agreement with simulated results.

\begin{figure}
    \centering
    \includegraphics[width=0.7\textwidth]{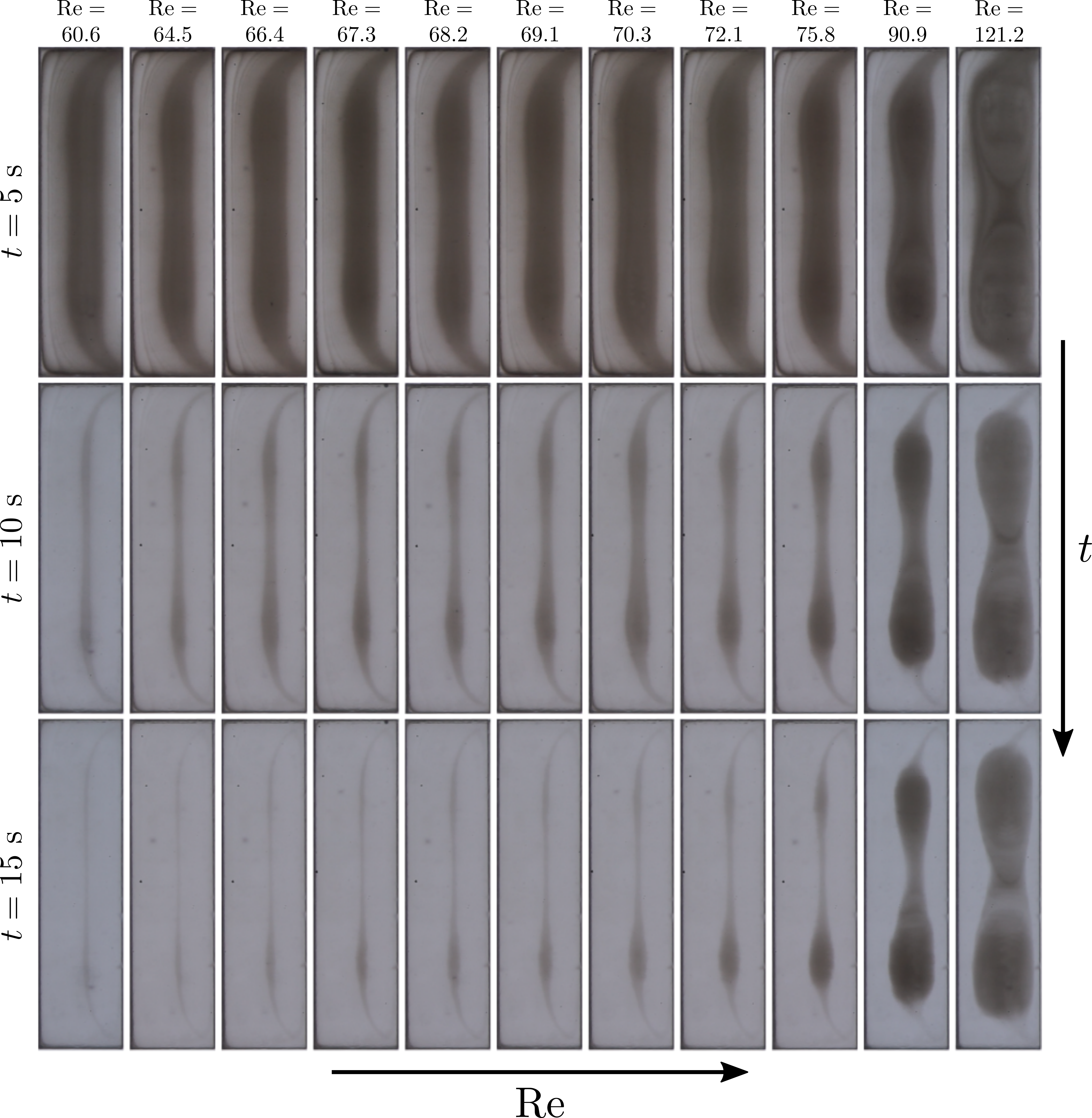}
    \caption{Time lapse and $\Re$ comparison of dye trapping in a channel with $w/\ell=3.96$. Results illustrate the transition from a case with no apparent dye trapping at $\Re=60.6$, where the only remaining dye at $t=15$ s is on the vortex core and is slowly diffusing away, to a case with significant dye trapping at $\Re=121.2$. These results also highlight the difficulty in identifying precise $\Re_\text{crit}$ transition numbers experimentally, since cases with very small vortex breakdown regions can appear indistinguishable from those without vortex breakdown. Similarly, this technique also has some uncertainty in identifying whether one or two vortex breakdown regions exist in the flow for different $\Re$. The flow is from left to right.}
    \label{fig:width_4_experimental}
\end{figure}

Due to the diffusive nature of the dye, identifying the exact transition Reynolds numbers precisely is difficult. This is because as the Reynolds number approaches the critical value for transition, the region in the vicinity of where the bubble will appear is already becoming a stagnation region, leading to long dye-flushing times even below $\recrit$. These cases are difficult to differentiate between those in which dye is trapped and slowly diffusing out of the breakdown bubble. Figure \ref{fig:width_4_experimental}
illustrates some of the difficulties in using this approach to precisely identify the transition Reynolds number $\recrit$. Here, there is an obvious distinction between the dynamics of the dye between the $\Re=60.6$ and the $\Re=121.2$ cases. For the lower $\Re$ case, at $t=15$ s, the only remaining dye is along the vortex core, which is a low velocity region. Sufficiently long time is needed before the dye can diffuse off of he vortex core and finally be flushed downstream. In contrast, for the higher $\Re$ case, a significant amount of dye is clearly trapped in some recirculating flow structure.

Hence, identifying the precise transition between these two behaviors is not trivial, especially because the vortex breakdown regions shrink smoothly and eventually vanish as $\Re$ is decreased. Thus, near the transition $\Re$ values, the vortex breakdown regions are typically very small, such that the dye can diffuse out of them relatively quickly, and the dye-trapping visualizations may appear nearly identical to those slightly below the transition $\Re$. Thus, this effect introduces some uncertainty into the experimental measurements of $\Re_\text{crit}$. Nonetheless, the results agree reasonably well with numerical simulations for most cases. For ambiguous cases, we simply label those experimental results as transition cases without specifying whether they exhibit breakdown or not.

\section{Results and Discussion}\label{section:numerical}

\begin{figure}
\centering
\includegraphics[width=0.6\textwidth]{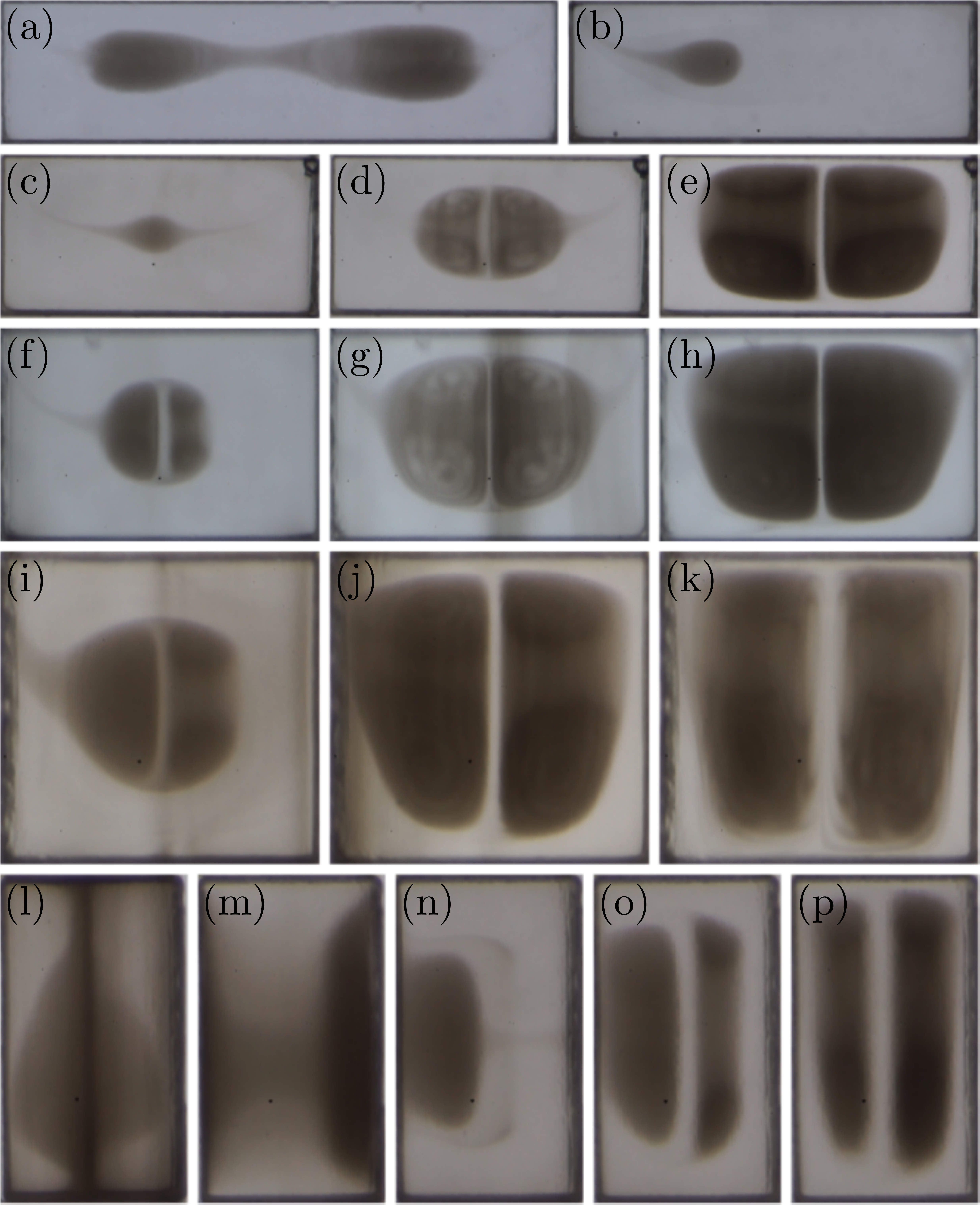}
\caption{Overview of the various types of dye trapping behavior observed in the microfluidic experiments over a range of cavity widths and Reynolds numbers. A variety of behaviors are observed including multiple separated trapping regions (e.g., (a)), individual symmetric (e.g., (c), (d)) or asymmetric (e.g., (b), (f)) trapping regions, large trapping regions that occupy nearly the entire cavity (e.g., (k)), and cases where dye appears to be trapped near the side walls (e.g., (m)). The values of $(w/\ell,\Re)$ for panels (a)-(p) are (3.96,91.0), (2.84,70.0), (1.97,41.1), (1.97,59.6), (1.97,119), (1.48,51.4), (1.48,77.9), (1.48,117), (1.01,56.2), (1.01,112), (1.01,225), (0.55,38.3), (0.55,53.2), (0.55,78.7), (0.55,106), and (0.55,213), respectively. The flow direction is from bottom to top.}
\label{fig:breakdown_examples}
\end{figure}

In this section, we present and discuss the results of our 3D numerical simulations and experiments. As a brief introduction to the types of vortex breakdown and trapping that have been observed in such systems, Figure \ref{fig:breakdown_examples}
presents characteristic results from the microfluidic experiments spanning a wide range of Reynolds numbers and cavity widths. These results illustrate a variety of observed behaviors including results that show two separated dye trapping regions as in panel (a), single asymmetric regions as in (b) or (f), single symmetric regions as in (d), large trapping regions that occupy nearly the entire cavity except near the walls as in (j), (k), or (p), and cases where the dye appears to be trapped near the side walls, as opposed to an internal bubble region as in (m). The internal structure of the recirculation regions can even be observed as in (d) and (g), where the swirling dye clearly illustrates the symmetric counter-rotating vortical structures inside the trapping regions. These preliminary observations indicate the types of behaviors we can expect to observe in the single-phase simulations, and they are used to inform the parameter regimes that we explore numerically.

\begin{figure}
\centering
\includegraphics[width=\linewidth]{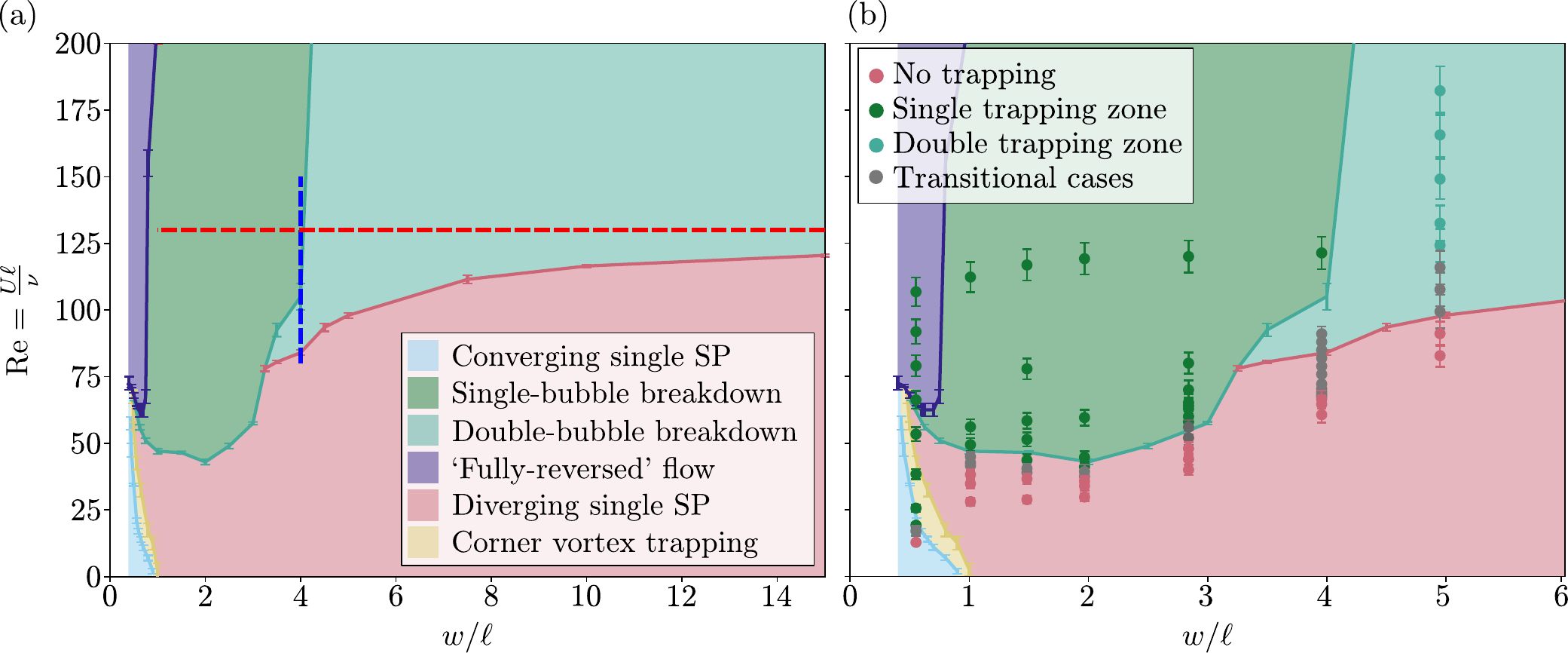}
\caption{Overview of the numerical and experimental results. Background colors indicate the different observed vortex behaviors which we have designated as: (1) converging single stagnation point (SP) mode, (2) single-bubble breakdown mode, (3) double-bubble breakdown mode, (4) `fully reversed' flow, (5) diverging single stagnation point mode, and (6) corner vortex trapping. Solid lines indicate the critical Reynolds numbers identified in the simulations for transitions between modes. Panel (a) shows just the numerical results over the full range of $w/\ell$, and panel (b) includes the experimental results as data points overlayed on the phase diagram. The dashed blue and red lines in panel (a) correspond to the streamline visualizations presented in Figures \ref{fig:012bubble} and \ref{fig:varywidth}, respectively.}
\label{fig:re_v_w}
\end{figure}

Through simulating the shear-driven cavity flow across Reynolds numbers ranging from $\Re=0$ to 450 and cavity widths ranging from approximately $w/\ell=0.4$ to 15, we have identified six different flow regimes based on the vortex dynamics in the cavity, reflecting the complex dynamical transitions possible in such systems. To present all of these transitions and flow regimes effectively, we first present the full phase diagram capturing all of the transitions, and then describe the individual modes and transitions in detail. The phase diagrams summarizing the experimental and numerical results are shown in Figure \ref{fig:re_v_w}.
In this figure, the background colors indicate the different modes of the observed vortex behaviors in the cavity, which we designate as: (1) converging single stagnation point (SP) mode, (2) single-bubble breakdown mode, (3) double-bubble breakdown mode, (4) `fully reversed' flow, (5) diverging single stagnation point mode, and (6) corner vortex trapping. Each of these modes will be described and explored below. The solid lines in the figure represent the numerically calculated critical Reynolds numbers at which transitions occur between different modes. Panel (a) shows the results over the full range of simulated $w/\ell$, while panel (b) restricts the view to the regime in which experiments were performed, and the experimental results (symbols) are overlayed on the phase diagram. The colors of the symbols denote whether the experiments demonstrated no trapping of dye, trapping of dye in a single region, trapping of dye in two separate regions, or transitional cases that are difficult to categorize just by visual inspection. The dashed blue and red lines in panel (a) correspond to the streamline visualizations presented in Figures \ref{fig:012bubble} and \ref{fig:varywidth}, respectively. As indicated by the phase diagram, the shear-driven flow in a cavity experiences a complex landscape of dynamical transitions. In Section \ref{sec:large_widths}, we investigate the large cavity width regime corresponding to approximately $w/\ell>4.5$ in which only a single transition is observed. Next, in Section \ref{sec:medium_widths}, we consider intermediate cavity widths in the range $1\leq w/\ell\leq 4.5$ in which an additional transition type appears for higher critical Reynolds numbers. Finally, in Section \ref{sec:small_widths}, we explore the relatively narrow width cavities with widths of approximately $w/\ell<1$ in which several additional vortex modes and dynamical transitions appear.

\subsection{Large cavity widths ($\mathbf{w/\boldsymbol{\ell}\geq4.5}$)}\label{sec:large_widths}
In the limit of very large cavity widths, the bulk of the flow in the cavity behaves as approximately 2D away from the sidewalls. Therefore, if any vortex breakdown is to occur in such cases, it must be localized near the side walls, where a flow component along the vortex axis is significant. Thus, if vortex breakdown is to occur in such systems, it must consist of two distinct separated bubble breakdown regions, which should be relatively insensitive to $w/\ell$ since they are well-separated. This is precisely what we observe in the shear-driven cavity flow at large $w/\ell$. As shown in Figure \ref{fig:re_v_w}, only two possible modes exist above $w/\ell\approx4.5$, consisting of the modes we have designated as the ``diverging single stagnation point'' mode and the ``double-bubble breakdown'' mode. Furthermore, as $w/\ell\rightarrow15$, the critical Reynolds number transition between these two modes asymptotes to $\Re_\text{crit}\approx125$, independent of $w/\ell$.

\begin{figure}
\centering
\includegraphics[width=0.7\textwidth]{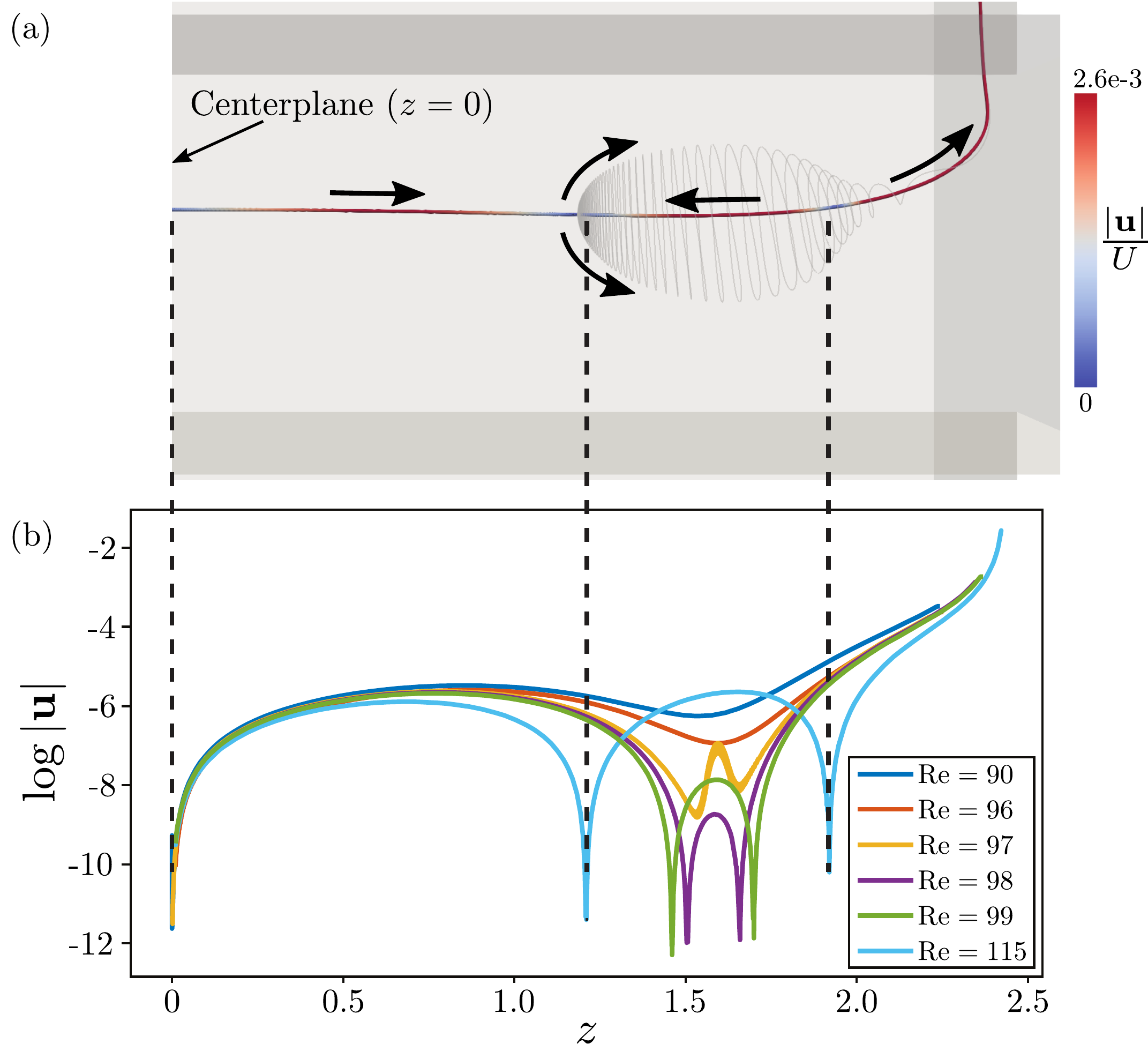}
\caption{The formation of the ``double-bubble breakdown'' mode as the Reynolds number is increased across the gray transition line in Figure \ref{fig:re_v_w}. Panel (a) shows the qualitative structure of the double-bubble breakdown mode in which low-velocity flow proceeds along the vortex axis, decelerates to an internal stagnation point, diverges along the surface of the breakdown bubble, converges to a second stagnation point, and then diverges both downstream toward the outlet and into the breakdown bubble where it recirculates. Here $w/\ell=5$ and  $\Re=115)$. The vortex core is colored by the velocity magnitude, highlighting the stagnation points (blue) in the flow. Panel (b) shows the magnitude of the velocity along the vortex core and highlights the appearance of the stagnation points as the Reynolds number approaches and crosses the critical value. As can be seen, as the Reynolds number approaches the critical value, a low-velocity region begins to appear, which splits into two low-velocity regions around $\Re=97$, finally becoming true stagnation points around $\Re=98$ and demarcating the boundaries of the breakdown bubble.}
\label{fig:stagpointemergence}
\end{figure}

The first of these two modes, the ``diverging single stagnation point'' mode is the same as the lower $\Re$ example described by \citeauthor{torczynski} \cite{torczynski}, which is shown diagrammatically in Figure \ref{fig:torczdemo}a and as a streamline visualization in Figure \ref{fig:torczdemo}b. In this mode, the flow on the channel centerplane $z=0$ enters the cavity, spirals down towards an internal stagnation point (corresponding to the vortex center in the 2D and nearly 2D cases), then diverges toward both side walls along the vortex core before ultimately exiting the cavity and proceeding downstream. Generally, the axial velocity component on the vortex core is extremely small, especially for large $w/\ell$, except near the side walls where the flow becomes fully 3D. Thus, the variation of the relative magnitude of axial and swirl flow along the vortex core with respect to the axial position and Reynolds number sets up the circumstances where vortex breakdown may occur. This is precisely what we see as we increase the Reynolds number above the critical solid gray transition line in Figure \ref{fig:re_v_w}. Specifically, two distinct vortex breakdown bubbles form simultaneously as this limit is crossed, creating the ``double-bubble breakdown'' mode. The appearance of this mode is visualized in Figure \ref{fig:stagpointemergence}.
Here, panel (a) first visualizes the form of the double-bubble breakdown mode. The vortex core is colored by the magnitude of the flow velocity, which highlights the stagnation points (blue regions) on the channel centerplane as well as bounding the upstream and downstream limits of the breakdown bubble. As can be seen, flow along the vortex axis proceeds towards the side wall, decelerating towards an internal stagnation point. The flow then diverges out along the streamsurface that bounds the bubble shape (see the grey streamline in Figure \ref{fig:stagpointemergence}(a)), converging at the back towards a second internal stagnation point. Here, the flow again diverges either out of the cavity towards the outlet or into the breakdown bubble where the flow continues to recirculate. Note that the flow is symmetric about the channel centerplane $z=0$ such that the second vortex breakdown region exists in the proximity of the other side wall. Panel (b) illustrates how these vortex breakdown regions originally form as the critical Reynolds number transition line is crossed. Here, the curves indicate the magnitude of the fluid velocity on the vortex core as a function of axial position and $\Re$. In this case, $\Re_\text{crit}\approx98$. As can be seen, even below $\Re_\text{crit}$, the fluid velocity in the vicinity of the breakdown region is relatively low. As $\Re\rightarrow\Re_\text{crit}$, the velocity magnitude begins to drop in the location where the breakdown will occur. At $\Re=97$, this low-velocity region splits into two even lower-velocity regions. Finally, at $\Re\approx98$, two distinct stagnation points appear. This indicates the onset of vortex breakdown and demarcates the extent of the breakdown bubble. As $\Re$ increases beyond $\Re_\text{crit}$, the stagnation points separate farther,, and the vortex breakdown region enlarges.

\begin{figure}
\centering
\includegraphics[width=0.9\textwidth]{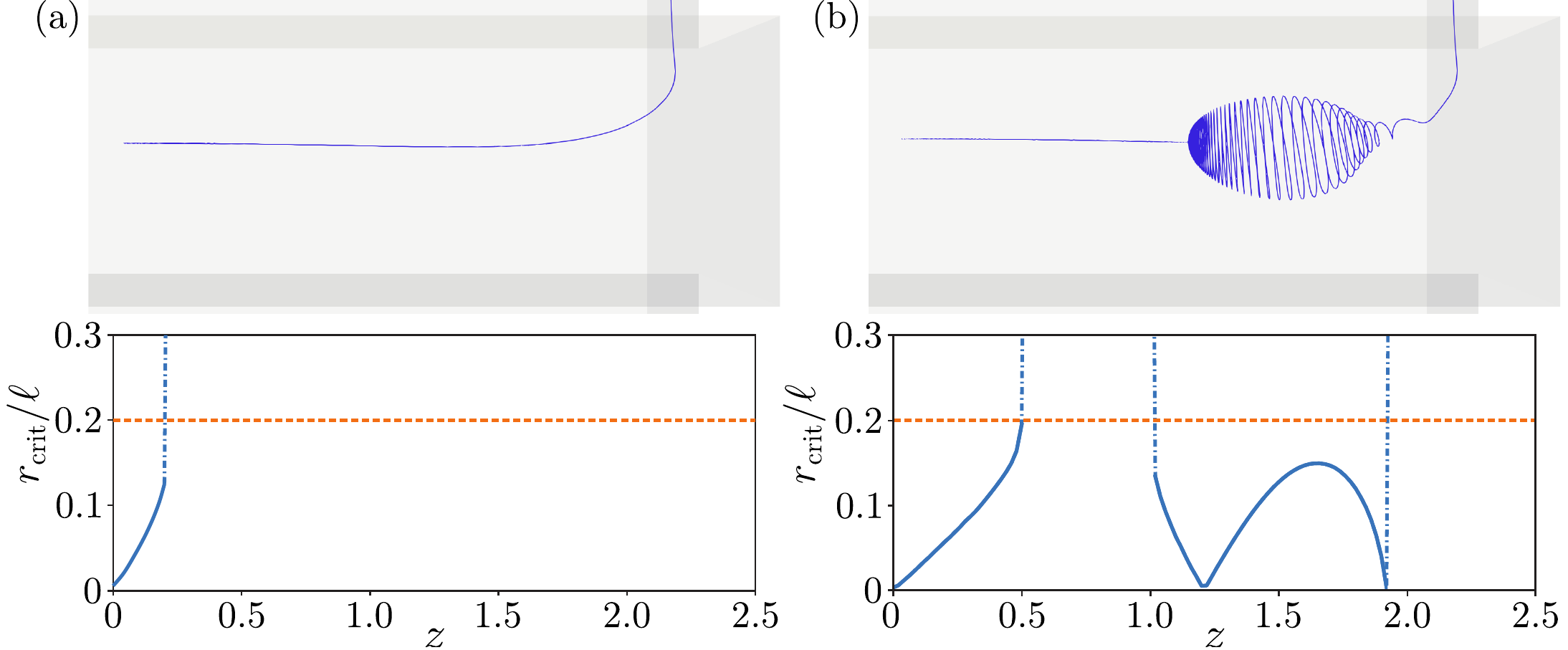}
\caption{Numerically computed $r_\text{crit}$ values as outlined by \citet{benjamin1962} for one half of the symmetric flow in a cavity of width $w/\ell=5$ at Reynolds numbers of (a) $\Re=90$ and (b) $\Re=115$. At $\Re=90$, no recirculation is evident in the streamline visualization, and no roots are found to the solutions of Equation (\ref{eq:criticality}) within the domain, except very near the centerplane, where the flow is approximately 2D. In contrast, at $\Re=115$, a clear recirculation region is seen in the streamline visualization, which corresponds to a pocket of subcriticality according to the roots of solutions to Equation (\ref{eq:criticality}). The dashed blue lines indicate regions where there either are no roots to the solutions of Equation (\ref{eq:criticality}) within the domain, corresponding to supercritical regions. The threshold radius of $r_{\text{crit}}=\ell/5$ was chosen based on the sampling radius within which the flow was axisymmetrized.}\label{fig:criticality}
\end{figure}

Before proceeding to consider the additional modes and transitions that occur for smaller cavity widths, we provide additional evidence that the recirculating flow features we observe are indeed vortex breakdown. We calculate the criticality condition as described in Section \ref{section:numerical_methods} to determine whether these recirculating flows satisfy the vortex breakdown interpretation of \citet{benjamin1962}. For illustration, we calculate the critical radius $r_\text{crit}$ in one half of the symmetric cavity for the cases $\Re=90$ and $\Re=115$ at a cavity width of $w/\ell=5$. The first of these is an example of the diverging single stagnation point mode, and the second is an example of the double-bubble breakdown mode. The calculated $r_\text{crit}$ values as a function of axial position $z$ for these two cases are shown in Figures \ref{fig:criticality}a
and \ref{fig:criticality}b, respectively, along with streamline visualizations that originate on the vortex core for comparison. Recall from Section \ref{section:numerical_methods} that if the solution to Equation (\ref{eq:criticality}) has a root at a location $r_\text{crit}$ within some threshold of the domain, then the region is subcritical according to Benjamin's criterion. Otherwise, if no such root exists within that threshold, then the flow is supercritical. As can be seen in Figure \ref{fig:criticality}(a), at $\Re=90$ there are no roots to the solution of Equation (\ref{eq:criticality}) along the vortex core except near the centerplane where the flow is essentially 2D. However, at $\Re=115$ (see panel (b)), a region of subcriticality clearly exists that coincides with the location of the recirculation bubble in the streamline visualization. Furthermore, the locations where $r_\text{crit}$ becomes very small also appear to coincide with the location of the internal stagnation points that bound the bubble. Thus, the recirculation regions we observe apparently satisfy the criticality condition developed by Benjamin for axisymmetric vortex breakdown. For additional insights and examples of these ideas, see e.g., Refs. \cite{ault2016vortex, chan2019coupling, ruith2003three}.

\subsection{Intermediate cavity widths ($\mathbf{1 \le w/\boldsymbol{\ell} \le 4.5})$}\label{sec:medium_widths}

As described in the previous section, for large cavity widths only two distinct vortex breakdown modes exist, with a single transition line between them. At smaller cavity widths of approximately $1\leq w/\ell \leq 4.5$, the situation becomes slightly more complicated, as shown in Figure \ref{fig:re_v_w}. Here, the flow can exhibit a third mode, the ``single-bubble breakdown'' mode, associated with two additional types of transitions. These correspond to the transition from diverging single stagnation point mode to the single-bubble breakdown mode, as wells as the transition between double-bubble breakdown mode and single-bubble breakdown mode. The single-bubble breakdown mode corresponds to that originally observed by \citeauthor{torczynski} \cite{torczynski} for a channel of width $w/\ell=2.125$ (see Figure \ref{fig:torczdemo}c and \ref{fig:torczdemo}d). In this mode, as in the diverging single stagnation point mode, the flow on the centerplane enters the cavity and proceeds to spiral down towards the vortex core. However, rather than spiraling down to a stagnation point, the flow in the single-bubble breakdown mode spirals down to a limit cycle. The flow then diverges off of the centerplane along the streamsurface of the recirculation bubble and spirals down to two bounding stagnation points. Here, the flow again diverges both downstream out of the cavity and toward the outlet, as well as into the recirculation zone along the vortex axis to the centerplane stagnation point. From here, the flow diverges off of the vortex axis on the centerplane and spirals out toward the limit cycle, completing the recirculation. Having described the structural features of this mode, we now consider transitions that occur in the moderate cavity width regime.

First, when the cavity width is approximately within the range $1\leq w/\ell \leq 3.5$, only a single transition can occur, namely the transition from the diverging single stagnation point mode to the single-bubble breakdown mode. This is precisely the transition found by \citeauthor{torczynski} \cite{torczynski}. As $\Re$ approaches and exceeds $\Re_\text{crit}$, the entirety of the single-bubble breakdown mode emerges from the original stagnation point of the diverging single stagnation point mode. When $\Re$ is just slightly above $\Re_\text{crit}$, the vortex breakdown bubble is quite small, and it grows in size as $\Re$ is increased. This can be seen clearly in Figures \ref{fig:breakdown_examples}c, \ref{fig:breakdown_examples}d, and \ref{fig:breakdown_examples}e, which correspond to $w/\ell=1.967$ and $\Re=41.1$, 59.6, and 119.2, respectively. For cavity widths approximately within the range $3.5\leq w/\ell\leq 4.5$, two transition lines are observed in Figure \ref{fig:re_v_w}. First, at the lower $\Re_\text{crit}$ value, the flow goes through the transition from the diverging single stagnation point mode to the double-bubble breakdown mode as described above for large cavity widths. Recall that as the Reynolds number increases for the double-bubble breakdown mode, the separation increases between the stagnation points bounding the breakdown region as the region grows. At the intermediate cavity widths between approximately 3.5 and 4.5, the separate breakdown regions are close enough to each other and the centerplane that their inner stagnation points can eventually merge with the original centerplane stagnation point, corresponding with a transition from the double-bubble breakdown mode to the single-bubble breakdown mode. When this occurs, the three stagnation points merge and become a single axially converging stagnation point and the limit cycle that was previously described. The shift in stability of the centerline stagnation point can be described as follows. Before the transition occurs, the centerplane stagnation point is stable to perturbations in the centerplane and unstable to perturbations on the vortex core, whereas after the stagnation points have merged, the centerline stagnation point shifts to be stable on the vortex core and unstable on the centerplane. The emergent limit cycle is stable on the centerplane and unstable along the surface of the breakdown bubble. This transition is analogous to that which was observed for the flow in a branching junction in which multiple vortex breakdown regions merged and underwent similar transitions in stability \cite{ault2016vortex}.

\begin{figure}
\centering
\includegraphics[width=\textwidth]{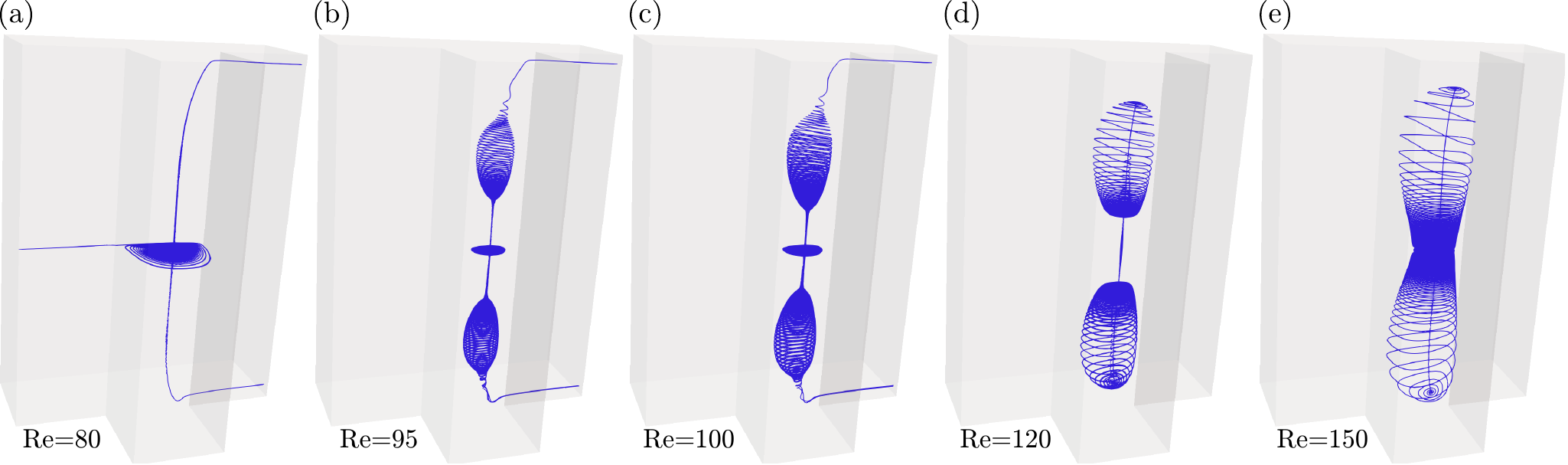}
\caption{Streamline visualization of the transition from diverging single stagnation point flow to double-bubble breakdown, and the subsequent transition to single-bubble breakdown as the Reynolds number is increased. Results correspond to a cavity width of $w/\ell=4$ with Reynolds numbers of $\Re=80$, 95, 100, 120, and 150 for panels (a) through (e), respectively. Results correspond to the blue dashed line in Figure \ref{fig:re_v_w}. Here, the critical Reynolds numbers for the two transitions are approximately $84\pm1$ and $105\pm 5$.}
\label{fig:012bubble}
\end{figure}

\begin{figure}
\centering
\includegraphics[width=\textwidth]{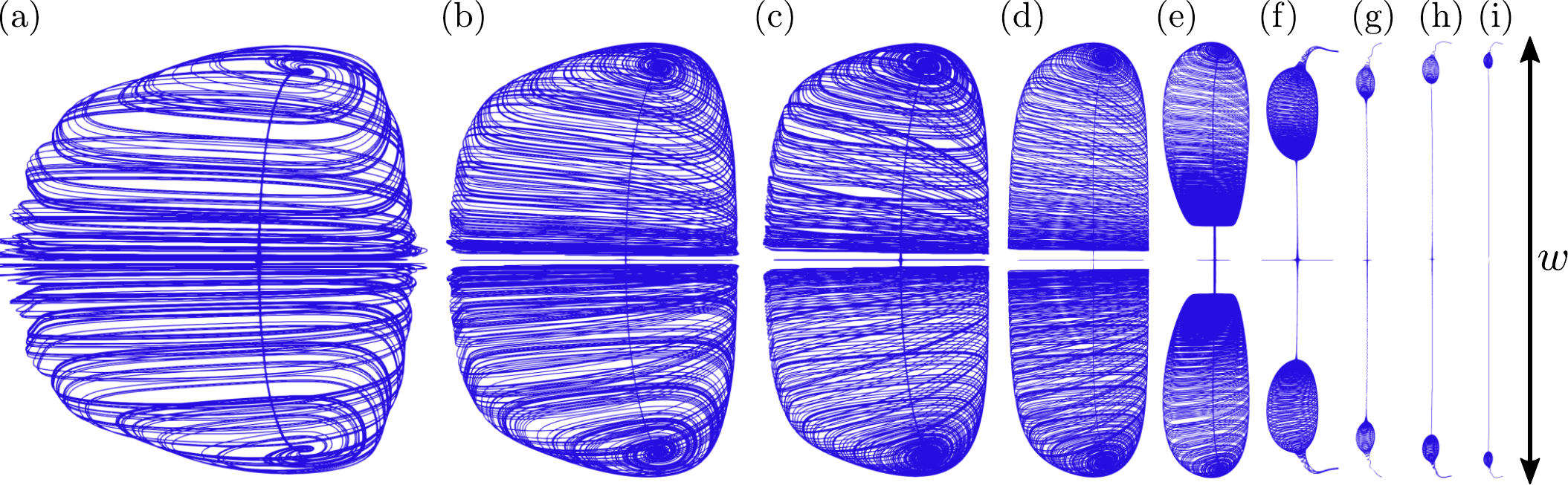}
\caption{Streamline visualizations of the transition from the single-bubble breakdown mode to the double-bubble breakdown mode as the cavity width increases. Here, the results correspond to a Reynolds number of $\Re=130$ for cavity widths of $w/\ell=1$, 1.5, 2, 3, 4, 5, 7.5, 10, and 15 for panels (a) through (i), respectively. Results are all rescaled to have the same width $w$ as shown in the scalebar.}
\label{fig:varywidth}
\end{figure}

The full transitioning process from diverging single stagnation point flow to double-bubble breakdown to single-bubble breakdown is shown in Figure \ref{fig:012bubble}
for a cavity of width $w/\ell=4$. Panels (a) through (e) correspond to Reynolds numbers of 80, 95, 100, 120, and 150, respectively, which correspond to the blue dashed line in Figure \ref{fig:re_v_w}a. Note that in Figure \ref{fig:re_v_w}, the transition between the double-bubble and single-bubble breakdown modes occurs at an $\Re_\text{crit}$ value that increases abruptly with channel width, especially when $w/\ell>4$. $\Re_\text{crit}$ for this transition increases from approximately 105 at a cavity width of $w/\ell=4$ to approximately 450 at a cavity width of $w/\ell=5$. It is possible that this transition still occurs at larger widths, although we did not simulate higher Reynolds numbers beyond 450 to find it. As the cavity width increases, the flow must inevitably reach a condition where it transitions to unsteady flow before the onset of this transition. So practically, the double-bubble to single-bubble transition is only likely to be observed for approximately $3.5\leq w/\ell \leq 4.5$. Finally, one more approach for visualizing the transition between the single-bubble and double-bubble breakdown is to hold the Reynolds number fixed and increase the cavity width. Streamline visualizations of these results at $\Re=130$ for cavity widths ranging from $w/\ell=1$ to 15 are shown in Figure \ref{fig:varywidth}.
At this Reynolds number, the width at which the transition occurs is approximately $w/\ell=4$. As $w/\ell$ increases in panels (f) through (i), the shape of the vortex breakdown regions become roughly independent of cavity width as they become well separated. Here, the results are shown side-by-side all rescaled to have the same width $w$. This is why the vortex breakdown regions appear to get smaller for larger $w/\ell$, because effectively $\ell$ is decreasing. However, if the results were rescaled to have the same $\ell$ dimension, the vortex breakdown regions in panels (g), (h), and (i) would look nearly identical with the same size, shape, and distance from the side wall.

\subsection{Narrow cavity widths ($\mathbf{w/\boldsymbol{\ell}<1})$}\label{sec:small_widths}

Finally, we consider the vortex dynamics and transitions in narrow cavity widths in the range $w/\ell<1$. As can be seen in Figure \ref{fig:re_v_w}, the narrow cavity case exhibits the most complex behaviors, with three new vortex modes unobserved at large cavity widths, and demonstrating a total of five possible modes and four possible dynamical transitions. First, note that our results and phase diagram only show results for $w/\ell>0.4$. In all of our simulations at cavity widths below 0.4, we never identified any structures that resembled vortex breakdown. Indeed, the 3D nature of the flow appears to change in this regime. We believe this is due to a transition into the Brinkman regime, where the flow essentially becomes a 3D lubrication type flow. That is, in the so-called Brinkman limit, when a flow is tightly confined in the depth direction, it can be treated as 2D in the depth-averaged sense \cite{leal2007advanced}. More precisely, the fluid velocity in such a limit is approximately only in the $x-$ and $y-$directions, with the magnitude being an approximately parabolic function of $z$, at least away from the walls. In this regime, the flow does still recirculate and remain trapped in the cavity, but only in the same manner as the truly 2D flow in a cavity, which is clearly not a form of vortex breakdown, since the velocity component along the vortex core is approximately zero. Thus, we consider the flow regime with $w/\ell<0.4$ to be a Brinkman flow regime, and outside the scope of this study.

\begin{figure}
\centering
\includegraphics[width=\textwidth]{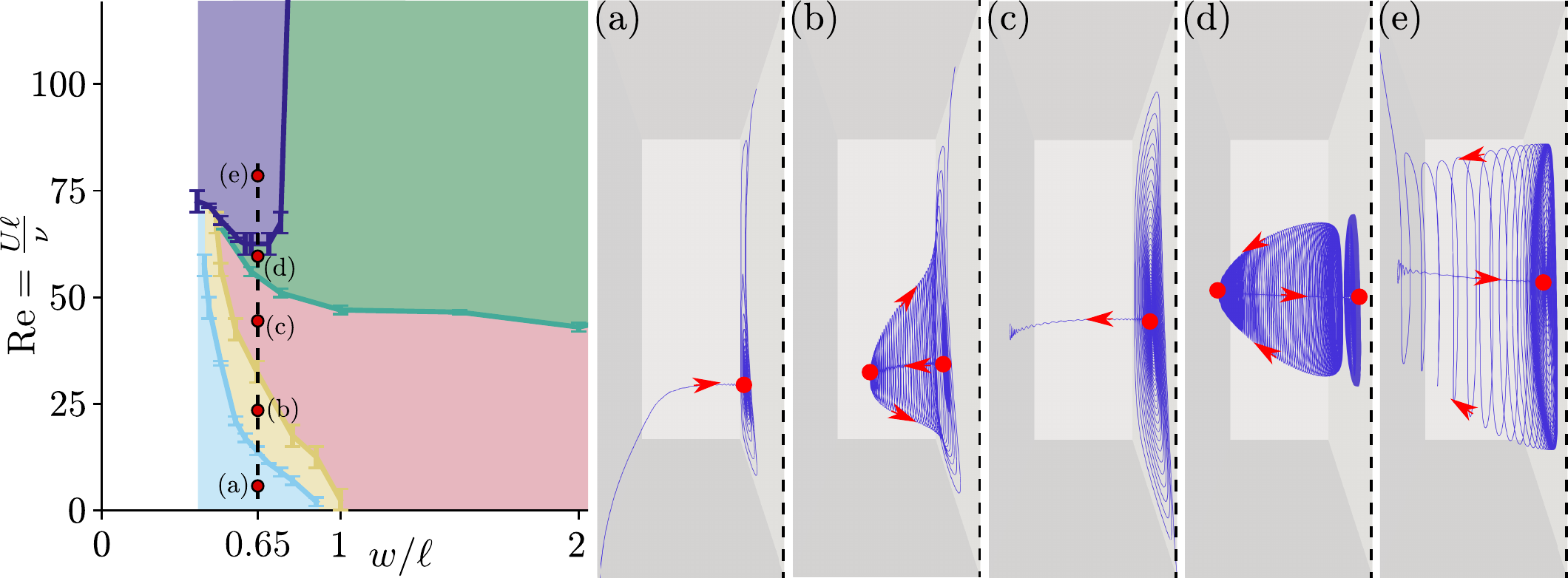}
\caption{Overview of the flow regimes that are present for low cavity widths of approximately $0.4< w/\ell<1$. Representative streamline visualizations are presented in panels (a) through (e) corresponding to each of the labeled red markers in the left panel. Background colors use the same legend as Figure \ref{fig:re_v_w}. Streamline visualizations correspond to $w/\ell=0.65$ and only show half of the channel, where the flow is understood to be symmetric about the channel centerplane which is on the right side of each panel. Flow directions are indicated on the streamlines, and the red dots in panels (a) through (e) represent the location of internal stagnation points.}\label{fig:lowW}
\end{figure}

The five different modes that are present over the range $0.4< w/\ell < 1$ are illustrated in Figure \ref{fig:lowW}.
Here, streamline visualizations are presented for each of the modes in panels (a) through (e) at a cavity width of $w/\ell=0.65$, corresponding to the five labeled red markers on the phase diagram. The background colors in the phase diagram match those used in Figure \ref{fig:re_v_w} and use the same legend. In the streamline visualizations, only half of the domain is shown, and the flow is symmetric about the channel centerplane, which is on the right side of each panel. Flow directions are highlighted by red arrows, and internal stagnation points are indicated by red circles. These five modes correspond to (a) converging single stagnation point mode, (b) corner vortex trapping, (c) diverging single stagnation point mode, (d) single-bubble breakdown mode, and (e) `fully reversed' flow. Thus, modes (c) and (d) are the same as those we have seen at higher cavity widths, but here they only exist over a much narrower range of Reynolds numbers.

Above the solid dark purple transition line, the single-bubble mode transitions into mode (e), which we have designated the `fully reversed' flow regime based on Harvey's \cite{harvey1962some} observation on the axial flow reversal in vortex breakdown. In their work, Harvey investigated the swirling flow in a cylindrical pipe as the magnitude of the swirl velocity was increased. When the magnitude of the swirl is low, the flow is a simple swirling pipe flow where the velocity along the vortex core points downstream everywhere. When the swirl is increased beyond a certain threshold, the flow exhibits bubble-type vortex breakdown, in which flow within the breakdown bubble is reversed and points upstream. As the magnitude of the swirl becomes even stronger, the breakdown bubble continues to expand and eventually engulfs the entire pipe domain, such that the velocity on the vortex core points upstream everywhere, and the stagnation point on the vortex core, which marks the beginning of the breakdown bubble, disappears. For this reason, Harvey classified this flow regime as `fully reversed'. This phenomenon is analogous to what we see in the shear-driven cavity flows for $0.4<w/\ell<1$. Starting from the diverging single stagnation point mode at point (c), the flow along the vortex core first proceeds `downstream', i.e., from the centerplane toward the side wall. Then, as the Reynolds number is increased to point (d), the flow transitions to single-bubble breakdown mode, where the flow inside the bubble is reversed and proceeds `upstream', toward the centerplane. Finally, increasing the Reynolds number to point (e), the vortex breakdown bubble grows until it engulfs the entire width of the cavity, and the flow everywhere along the vortex axis is `upstream' toward the channel centerplane.

\begin{figure}
\centering
\includegraphics[width=\textwidth]{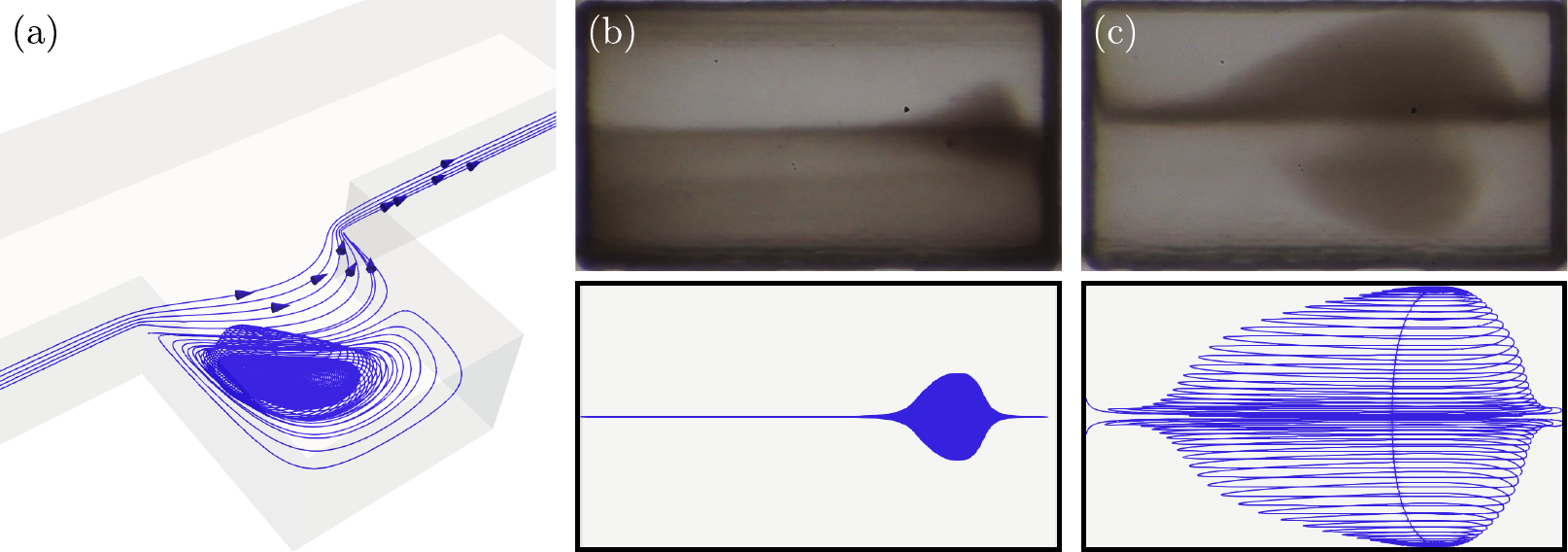}
\caption{Visualization of the corner vortex trapping mode corresponding to point (b) in Figure \ref{fig:lowW}. (a) Perspective view of the streamline visualizations of the recirculation bubble in a cavity of width $w/\ell=0.56$ at $\Re=30$. The recirculation bubble originates from the corner vortex that forms on the inlet side of the cavity. Panels (b) and (c) show bottom-side comparison views of the dye-trapping experiments and simulated streamlines at Reynolds numbers of (b) $\Re_\text{exp.}=19.2$ and $\Re_\text{sim.}=22$, and (c) $\Re_\text{exp.}=38.4$ and $\Re_\text{sim.}=40$.}\label{fig:CVT}
\end{figure}

One of the unique features of the narrow-width shear-driven flow in a cavity is the appearance of a second type of vortex breakdown trapping that can exist for surprisingly low Reynolds numbers. This corresponds to point (b) in Figure \ref{fig:lowW} and is the mode we have designated as `corner vortex trapping'. As can be seen in the phase diagram, this corner vortex trapping regime occupies a relatively small and narrow parameter space. Furthermore, this mode exhibits several unique features. First, the stability of the stagnation points is opposite to that of the single-bubble breakdown mode. In the single-bubble breakdown mode, the centerplane stagnation point is unstable on the centerplane and stable on the vortex core, whereas this is reversed for the centerplane stagnation point of the corner vortex trapping mode. Similarly, the stagnation point bounding the extend of the breakdown bubble is stable on the bubble streamsurface and unstable on the vortex axis for the single-bubble breakdown mode, but this is likewise reversed for the corner vortex trapping mode. Furthermore, a unique transition can be seen when going from points (c) to (b) in Figure \ref{fig:lowW}. Here, we begin with the diverging single-stagnation point mode, and by decreasing the Reynolds number a vortex breakdown bubble appears. This exhibits the opposite of the typically expected behavior in which the Reynolds number must be increased to lead to the onset of breakdown. At these low cavity widths and Reynolds numbers, the appearance of the flow past the leading corner of the cavity is analogous to a reattaching flow \cite{tritton1988physical}, and an eddy forms at the corner on the inlet side of the cavity. Thus, we have designated this new vortex trapping mode `corner vortex trapping' because it originates from this corner eddy in a slightly different location from the single-bubble breakdown mode. Several visualizations of this corner vortex trapping mode are shown in Figure \ref{fig:CVT}.
Here, panel (a) shows a perspective view of a streamline visualization of the trapping mode. As can be seen, the flow resembles a reattachment flow over a corner, and the recirculation bubble originates from the recirculating corner vortex on the inlet side of the cavity. Panels (b) and (c) show comparisons between the experimental results and simulated streamlines of the recirculation bubble at Reynolds numbers of approximately $\Re\approx20$ and $\Re\approx40$, respectively, highlighting the increase in size of this trapping mode with the Reynolds number.

Finally, the last mode which we have not discussed is the converging single stagnation point mode, corresponding to panel (a) in Figure \ref{fig:lowW}. Recall that in the diverging single stagnation point mode, the flow on the centerplane from the inlet spirals down toward a centerplane stagnation point that is stable on the plane and unstable on the vortex axis, and then diverges toward both side walls along the vortex axis and proceeds downstream to the outlet. Here, in the converging single stagnation point mode, the stability of the centerplane stagnation point is reversed, such that flow from both sides of the inlet spirals down to the vortex core, proceeds inwards toward the centerplane stagnation point, diverges off of the vortex core along the centerplane, spirals outward, and finally exits the cavity and proceeds to the outlet. The transition from this mode to the corner vortex trapping mode occurs as the solid purple $\Re_\text{crit}$ line is crossed in Figure \ref{fig:lowW}, and the bounding stagnation points and limit cycle emerge from the original centerplane stagnation point.

\subsection{Asymmetric breakdown}
In the previous sections, we have explored in detail the different vortex breakdown modes and flow regimes that were identified experimentally and computationally for the shear-driven flow in a cavity. One feature that was seen in the experimental dye-trapping visualizations but not observed in our initial numerical simulations was the presence of asymmetric trapping. For instance, see panels (b) and (n) in Figure \ref{fig:breakdown_examples}. Here, the dye-trapping is clearly asymmetric, and was most pronounced for cases with $3\lesssim w/\ell \lesssim 4$. For all of these cases, our simulations demonstrated symmetric trapping. However, it must be noted that the flow within the cavity is already a relatively low-velocity region relative to the outer channel flow. Further, along the vortex core, the axial velocity component is typically very small, especially when vortex breakdown occurs. Thus, it is possible that relatively small asymmetries in the outer channel flow could significantly bias the axial flow on the vortex core in the cavity, which would in turn significantly affect the structure of the vortex breakdown regions.

\begin{figure}
\centering
\includegraphics[width=0.7\textwidth]{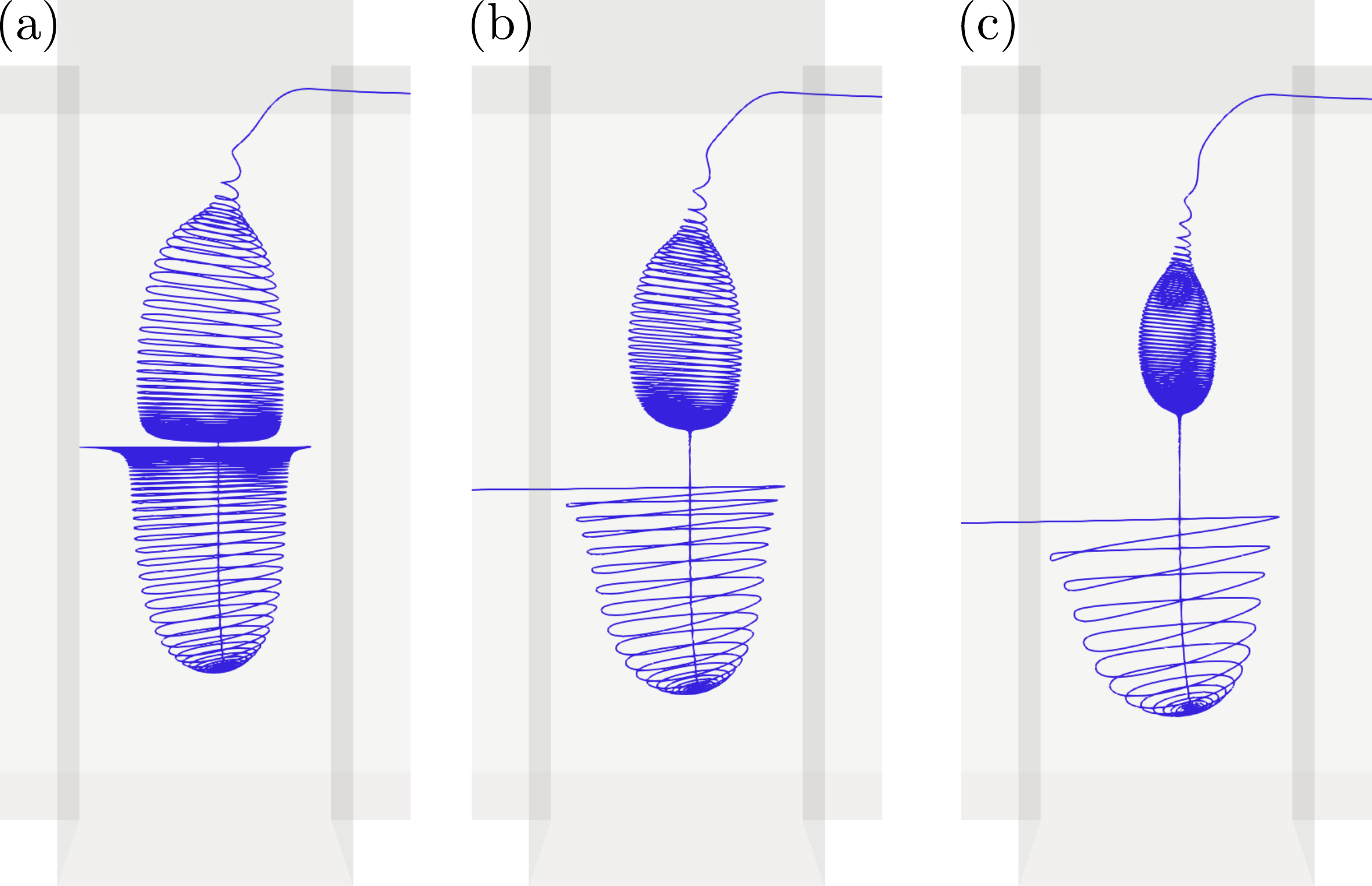}
\caption{Asymmetric vortex breakdown in a cavity with $w/\ell=3$ at $\Re=80$ with increasing degrees of asymmetry given by (a) $s=0.01$, (b) $s=0.1$, and (c) $s=0.2$. Here, the asymmetry in the inlet condition increases the velocity magnitude near the bottom sidewall of the channel and decreases it near the top sidewall.}\label{fig:asym_bubble}
\end{figure}

In order to test this hypothesis, we performed additional simulations with slightly asymmetric inlet conditions to test how sensitive the structure of the vortex breakdown regions is to any asymmetry in the outer channel flow. Here, we impose an inlet boundary condition on the simulations that consists of the typical fully developed inlet condition that was used in all other simulations multiplied by a constant linear function that varies from $1-s$ to $1+s$ from sidewall to sidewall, where $s$ represents the degree of asymmetry. Simulations were performed for a cavity of width $w/\ell=3$ at $\Re=80$ with values of $s=0.01$, 0.1 and 0.2. Since the flow is no long fully developed at the inlet, the flow in the cavity must also depend on the length of the inlet. Here, we keep the inlet length fixed as $2\ell$ and only vary~$s$. Streamline visualizations of these results are shown in Figure \ref{fig:asym_bubble}.
As can be seen, increasing $s$ clearly increases the asymmetry in the vortex breakdown dynamics in the cavity. As the degree of asymmetry increases, the obvious vortex breakdown bubble shifts to the side and begins to resemble one of the two bubbles present in the double-bubble breakdown mode. Furthermore, the single-bubble breakdown mode is apparently relatively sensitive to asymmetry. For example, even with $s=0.01$ as in Figure \ref{fig:asym_bubble}a, it is not clear whether the bottom half of the structure remains a recirculation zone, although it appears not to be. This is more obvious at larger values of $s$. Thus, a small asymmetry in the system could eliminate half of the trapping potential in the system.

\begin{figure}
\centering
\includegraphics[width=0.9\textwidth]{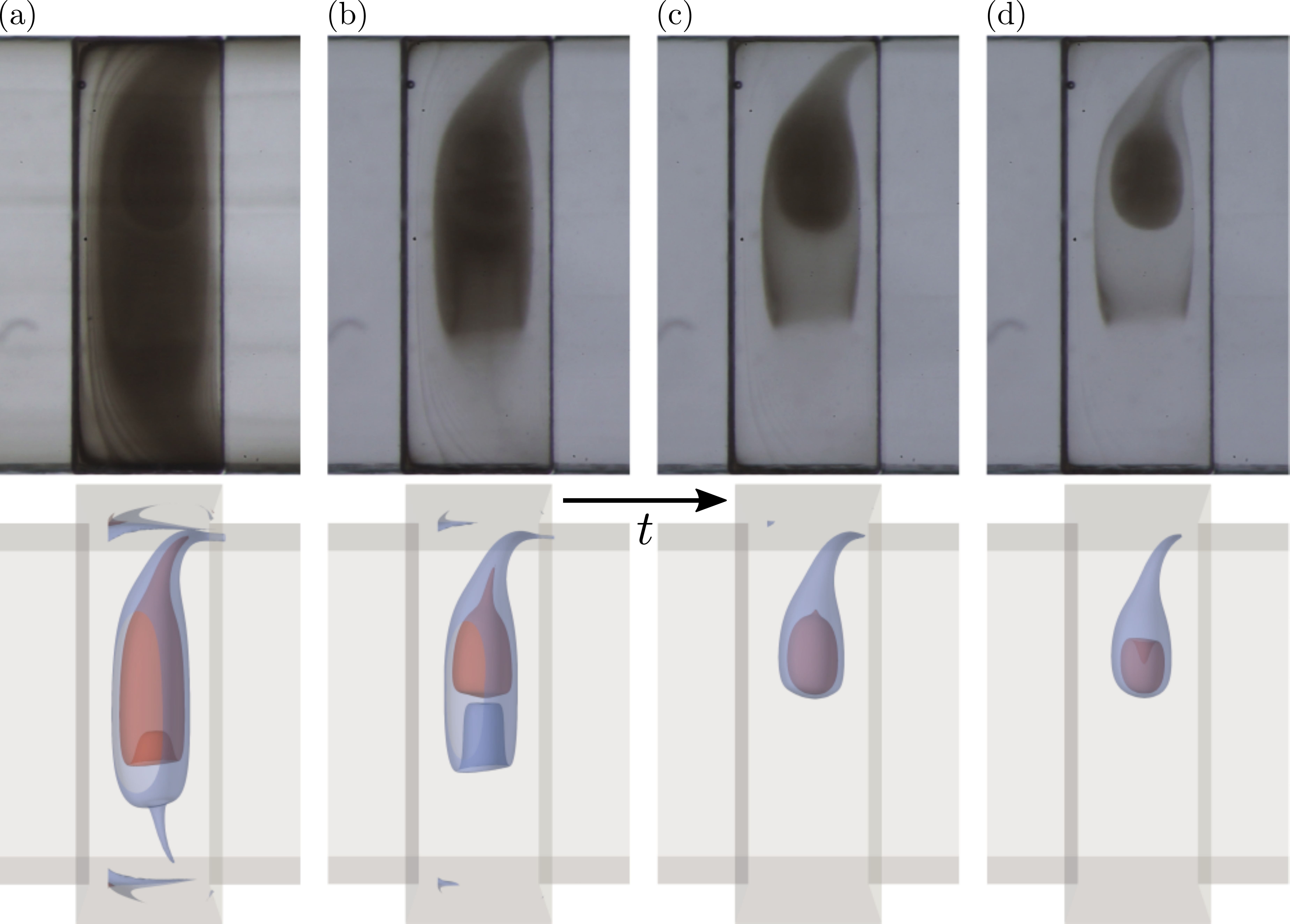}
\caption{Comparison between experimental results (top) at $w/\ell=2.84$, $\Re=80.0$ and numerical simulations (bottom) at $w/\ell=3$, $\Re=80$ with a degree of asymmetry of $s=0.1$. Results are compared at nondimensional times of (a) $t^*=360$, (b) $t^*=610$, (c) $t^*=810$, and (d) $t^*=1010$, where $t^*=tU/\ell$. Level sets of the concentration field are shown with $C=0.4C_0$ (blue) and $C=0.9C_0$ (red) to highlight the comparison.}\label{fig:asym_comp}
\end{figure}

In order to visualize the trapping computationally in these asymmetric cases, we used the \texttt{scalarTransportFoam} solver of OpenFOAM to simulate the advection-diffusion of a scalar concentration field $C$, representing the dye concentration in the experiments. Here, the initial condition for the concentration was set to a uniform value of $C_0$ everywhere. No-flux conditions were used at all of the walls, as well as at the outlet, and the concentration at the inlet was set to zero. Thus, these simulations are equivalent to taking the steady-state asymmetric velocity fields simulated above, filling the channel with a concentration field, and then flushing it away. A comparison between the experimental results and these simulations is shown in Figure \ref{fig:asym_comp}
at a cavity width of $w/\ell\approx3$ and $\Re=80$ with a degree of asymmetry given by $s=0.1$. Results are compared at nondimensional times of (a) $t^*=360$, (b) $t^*=610$, (c) $t^*=810$, and (d) $t^*=1010$, where $t$ has been nondimensionalized by the characteristic time for convection past the cavity, i.e., $t^*=tU/\ell$. Numerical results are shown as level sets of the concentration field with values of $C=0.4C_0$ (blue) and $C=0.9C_0$ (red) to highlight the comparison with the experimental results. Although we have used an artificial asymmetry in the inlet condition, which is unlikely to match that which was present in the experiments, we nonetheless see surprising similarities between the experimental and computational results. More detailed experiments with well-defined added asymmetries could be used to validate and expand on the sensitivity of these vortex breakdown features, which we leave as a topic for future work.

\section{Conclusion}\label{section:conclusions}
In this study, we have investigated the vortex breakdown and trapping mechanisms in the shear-driven flow in a rectangular cavity. Both fully 3D computational fluid dynamics simulations and microfluidics dye-trapping experiments were used to determine the different parameter regimes and vortex breakdown modes present in the system. A total of six main parameter regimes were identified, along with five different types of flow transitions along with the corresponding critical Reynolds numbers for each of these transitions as functions of cavity width. At large cavity widths of $w/\ell\geq 4.5$, only two modes were observed, the diverging single stagnation point mode and the double-bubble breakdown mode, along with the single transition between them. At smaller cavity widths of $1\leq w/\ell \leq 4.5$, the additional single-bubble breakdown mode appears, along with potential direct transitions to each of the other two modes. Streamline visualizations were used to highlight the dynamical changes that occur across these transitions and the consequences on the stability of stagnation points. We also used numerical simulations to show that the vortex breakdown features we observe satisfy the criticality condition presented by Benjamin in an axisymmetric swirling pipe flow. At smaller width of $w/\ell\leq 1$, the phase space becomes the most complicated, with a total of five possible modes and four possible transition types depending on $w/\ell$ and $\Re$. In particular, we see a new `fully reversed flow' mode at high Reynolds numbers, where the single-bubble breakdown mode has grown to engulf the entire width of the cavity. At lower Reynolds numbers, below those for diverging single stagnation point mode, we see the onset of the corner vortex trapping mode, which is an especially low-$\Re$ mode existing even down to $\Re=\mathcal{O}(1)$ and demonstrating recirculation that is in the opposite sense from that of single-bubble breakdown. Finally, we have considered the role of asymmetry in the channel or inlet condition as a possible source of the asymmetric dye-trapping seen experimentally at moderate cavity widths, and we showed that relatively small asymmetry in the inlet condition could have relatively significant effects on the structure of the vortex breakdown regions and the trapping capacity of the cavity. These results have provided new insights into the vortex breakdown and trapping dynamics in low- and moderate-Reynolds number flows in a cavity. These results highlight the importance of a detailed investigation into even basic flow geometries where the potential for vortex breakdown may be underappreciated, and the potential flow regimes and parameter spaces may be unexpectedly complicated.

\bibliography{references}
\end{document}